\documentclass[12pt]{article}
\usepackage{amsmath,amssymb,amsfonts,amsthm}
\usepackage{color}
\usepackage{multirow}
\usepackage{lscape}

\topmargin- 0.5cm
\newcounter{item}[section]
\newcounter{kirshr}
\newcounter{kirsha}
\newcounter{kirshb}
\newenvironment{mysect}[1]{\vskip8pt\par\noindent\setcounter{item}{1}

\setcounter{equation}{0}{\large\bf\arabic{section}.  #1 }\vskip8pt\nopagebreak\par\nopagebreak }
{\stepcounter{section}\upshape\par}

\newcommand\overcirc[1]{\raisebox{10pt}{\tiny$\circ$}{\kern-6.5pt}\mbox{$#1$}}
\newcommand\undersym[2]{\raisebox{-6pt}{\tiny$#2$}{\kern-5pt}\mbox{$#1$}}

\begin{document}

\title{\bf The Riddle of Gravitation}
\author{A. M.  Sid-Ahmed}
%\date{}
\maketitle                     % Produces the title.
\vspace{-1.13cm}
\begin{center}
{Department of Mathematics, Faculty of Science, Cairo
University}\end{center} \vspace{-0.8cm}
\begin{center}
{amrs@mailer.eun.eg }
\end{center}

\noindent{\bf Abstract.}
In the course of a decade, Einstein singlehandedly overthrew the
centuries-old Newtonian
framework and gave the world a radically new demonstrably
deeper understanding of gravity. It does not take much to get experts and nonexperts to
gush over the sheer brilliance and monumental originality of
Einstein's accomplishment in fashioning both the special and the general theories of relativity. There is no doubt that both theories capture the imagination.
The anti-intuitive properties of the
special theory of relativity and its deep philosophical implications, the bizzare and dazzling predictions of the general theory of relativity: the curvature of spacetime,
the exotic characteristics of black holes,
the bewildering prospects of gravitational waves, the discovery of astronomical objects as quasers and pulsers, the expansion and
the (possible) recontraction of the universe\ldots, are
all breathtaking phenomena. In this paper, we give a philosophical non-technical treatment of
both the special and the general theory of relativity together
with an exposition of some of the latest physical theories. We then give an outline of
an axiomatic approach to relativity theories due to Andreka and Nemeti\footnote{the Hungerian Professors at the mathematical
institute of Hungarian Academy of Science.} that throws light on the
logical structure of both theories. This is followed by an exposition of some of the bewildering
results established by Andreka and Nemeti concerning the foundations of mathematics
using the notion of relativistic computers.
We next give a survey on the
meaning and philosophical implications of the the quantum theory and
end the paper by an imaginary debate between Einstein and Neils Bohr
reflecting both Einstein's and Bohr's philosophical views on the quantum world.

\bigskip

The paper is written in a somewhat untraditional manner; there are too many footnotes. The reason behind
this is the following:  ignoring the footnotes, the paper is intended to be complete in itself.
The footnotes, on the
other hand, deal mainly
with the more technical
issues of both the special and general theories of relativity together with the intricate concepts
of the general theory of relativity and its connection to other modern physical theories.
In order not to burden the reader with
all the details, we have collected the more advanced material (mostly of philosophical nature) in the footnotes. We think that this
makes the paper easier to read and simpler to follow.
In fact, the paper without the footnotes should
be understood by anyone having a good scientific or a philosophical background. The paper in full is adressed
more to experts.

\begin{mysect}{Einstein, the Man and the Philosopher}

Most of the material presented  here may be found in \cite{AEI} and \cite{P}.

\bigskip

{\it Still there are moments when one feels free from one's own identification with human limitations and inadequacies. At such moments,
one imagines that one stands on some spot of a small planet, gazing in amazment at the cold yet profoundly moving beauty of the
eternal, the unfathomable; life and death flows into one, and there is neither evolution nor destiny; only being.} Einstein

\bigskip

{\it Joy in looking and comprehending is nature's most beautiful gift.} Einstein

\bigskip

{\it Nature in its simple truth is greater and more natural than any creation of human hands, than all the illusions
created by the spirit}. Robert Mayer as quoted in \cite{AEI}.

\bigskip

For more than two centuries, Newton's system had been regarded as the ultimate solution to the fundamental problems of science; as the final
and preordained picture of the world.\footnote{In his book, \lq\lq Principia Mathematica\rq\rq - the most influential scientific book of all times - Newton
was able to lay down the fundamental laws of mechanics in the language of differential calculus using the notion of rate of change. He then
succeeded to deduce Kepler's laws and generalize Galileo's ideas in his three famous laws of nature. Newton was thus the first to realize
that {\it the universe operated according to strict mathematical laws!}} This estimation is reflected in Alexander Pope's verse:

\bigskip

{\it Nature and Nature's law lay hid in night;}

{\it God said, let Newton be and all was light.}

\bigskip

Then came Einstein with his theory of relativity, and some wit added the lines:

\bigskip

{\it But not for long. Let Einstein be! the devil said;}

{\it And lo, it was dark: the light has fled.}

\bigskip

Albert Einstein, the most famous scientist of the twentieth century, perhaps the greatest genious of all time,
whose title to fame is uncontested and whose impact on modern scientific thought is unchallenged, was a man of touching simplicity and modesty.
He was always surprised by
the masses' infatuation with his theories, because they could not possibly mean anything to them. Those who
could understand his work did not crowd around him like primitives at the feet of their idol. They read, criticized and
discussed him.

\bigskip

Einstein never sought fame, fortune or glory. He did nothing to outclass others and certainly nothing to please others.
He strove to understand and resolve the laws of nature. This, he did, more like an amateur.  He was an original
and a deep thinker, who presented the world with a diamand mine. His ideas have led
directly to the era of atomic power and the dawning of the space age.
He can be described as the Copernicus of the twentieth century.

\bigskip

Einstein did not create his own legend as so many others believe they should.
When asked to give a portrait of himself towards the end of his life he wrote: \lq\lq Of what is significant in one's own existence
one is hardly aware, and it certainly should not bother the other fellow. What does a fish know about the water in
which he swims all his life.\rq\rq

\bigskip

Einstein was never a good pupil. His grades were mediocre and he never reached the top of his class. His family
thought he was retarded as he did not talk until the age of three. Strangly enough, what attracted him, apart from
mathematics and physics, was religious matters. His father, a far from orthodox Jew, sent him to a Catholic institution. There
he acquired a great interest in \lq\lq divinity\rq\rq, the biblical legends and the epic of Christ.

\bigskip

At the age of ten, he left primary school for the Lutpoid Gymnasium in Munich. Again, he showed no more brilliance than at
primary school. It was thanks to books of popular science, however, that the young Albert felt a passion for knowledge awake
in him. His favorite classical writers were Hume, Schiller and Goethe.

\bigskip

At an earlier age, his mother had made him take violin lessons. His rather mild enthusiasm was transformed gradually into
a passion for music. Music became his favourite past-time. From the age of fourteen,
he took part in domestic concerts. Mozart's
music played the same role in Einstein's life as Euclidean geometry did in his scientific development.

\bigskip

Einstein came up against a wall of refusals when he tried to obtain a post as an assistant on the completion of his studies. His
views were too original with an independent character that terrified his professors. He also had the misfortune to be born a Jew.
 Although officially there was no anti-Semitism in Switzerland, racial prejudice, as elsewhere, was not unknown. To earn a
living, Einstein was forced to accept a regular job as an engineer in the Patent Office in Bern. Einstein's life in Bern can be compared
with Newton's Woolsthorpe period during the plague of $1665-1667$, when he had to go away to Cambridge. It was in
Woolsthorpe that Newton developed his ideas on differential calculas, universal gravitation and the breaking down of light into
monochromatic rays. It was in Bern that Einstein developed the theory of Brownian motion, the photon theory (the photo-electric effect) and
the special theory of relativity.\footnote{In the current literature, the special theory of relativity is attributed to Lorentz, Poincare and Einstein. We
believe that this is not very fair. Though both Lorentz and Poincare contributed to the theory, it was Einstein who formulated
the theory on solid {\it physical} grounds, revolutionizing Newton's absolute space and time.}
Each one of these contributions could have earned him a Nobel price.\footnote{Einstein recieved his Nobel price
as a reward on \lq\lq
his work on the photo-electric effect and his contributions to theoretical physics\,\rq\rq. Both his special and
general theories of relativity were not yet properly understood!}

\bigskip

Also to be noted is Einstein's attitude towards philosophical literature. He ascribed purely aesthetic value to many
philosophical works. At the same time, he ascribed great philosophical and scientific value to works of fiction. His attitude was that of
a sympathetic listener accepting a philosophical point of view with an indulgent or ironical - as the case might be - smile.
His attitude
towards the 18-th and 19-th century philosophy may be summed in the following way: {\it Einstein's problem was whether or not it was possible
to deduce from observation of physical phenomena the causal relationship between them.} Hume's answer is negative: It is impossible to
penetrate into the causality of observable phenomena. Human understanding should be restricted to the phenomena themselves. However, Einstein's
conception of the real world of matter as the cause of sense impressions and of the cognisability of the objective laws of motion were not
in the least shaken by reading Hume. Einstein proceeds from the idea that a series of observable phenomena does not determine unequivocally
the nature of the causal relationships between them. Hence the picture of causal relationship is to some degree deduced independently of
direct observations. Einstein speaks of the free construction of concepts expressing causal relationships. Does this mean that such concepts
are {\it a priori} or conceptual, or that causal concepts are arbitrary as a whole? The answer is no. The causal connections of processes may
be expressed by means of different kinds of constructions. In this sense, their choice is arbitrary. But they must be in agreement with
observation; and it is our duty to select the construction which agrees best.

\bigskip

When Einstein was trying to formulate both his special and general theories of relativity\footnote{The theory of general relativity is based on a very simple idea,
which can be explained to a teenager. One must only imagine the
experience of falling and recall that those who fall have no sensation of weight. In the hands of Einstein, this everyday fact became
the openning to a profound shift in our way of understanding the world: while you can abolish the effects of
gravity locally, by freely falling, this can never be done over a large region of spacetime. Therefore,
while {\bf curved} space(time) can be approximated by a patchwork of small flat regions,
these regions will always have discontinuities where
we try to join them at their edges. This could be taken to mean that the overall space is curved. To put it in a more suggestive and provocative way:
{\it the very
fact of this failure to join smoothly is the curvature of space.}}, he did not study mathematics, nor did he supplement his scientific training by
looking at the latest experiments on the speed of light. Rather, he read about Ernst Mach, Immanual Kant and David Hume. His essential thinking was
philosophical, thinking deeply about the meaning of science, the problem of knowledge and the philosophical meaning of spacetime. In this way,
he was brought to ask some fundamental questions.\footnote{At the center of the discussion about quantum theory was a
great debate between Einstein and the great Danish physicist Neils Bohr\cite{EB}. Nearly every time they met, from their
first meeting until Einstein's death, they argued about the meaning of quantum theory. Einstein believed that the goal of physics was to
construct a description of the world as it would be in our absence. For Einstein, probabilistic theories were extremely interesting and
important and beautiful. But they were neither fundamental physical theories, nor objective. They were subjective theories, theories which we have to
introduce because of the fragmentary character of our knowledge. On the other hand, Bohr believed that this was impossible. For him, physics
was an extension of the common language, which people used to apprise each other of the results of their observations of nature. Einstein's
view of nature was classical, believing that it is possible to construct an objective picture of the world, and thereby capture
something of the eternal transcendent reality behind nature. Bohr, however, believed in the principle that properties are only defined relationally,
and that physics is an aspect of our relation with the world. Einstein revealed his uneasiness and discomfart
with the quantum theory in his two
famous sayings:\,
\lq\lq God does not play dice with nature \rq\rq and \lq\lq God is subtle but not malicious \rq\rq (\cite{EB}, \cite{Brian}.)
According to Einstein, he had spent hundred times more effort in trying to understand the quantum theory as opposed to his general theory of relativity.
A debate between Einsein and Bohr will be given at the end of the paper.}

\bigskip

Einstein always began with the simplest possible idea when confronting a problem, no matter how complicated.
Then, by describing how he saw the problem, he put it into the appropriate context.\,This intuitive
approach was like painting a picture. One could even speak of the music of his work; following Einstein's
train of thought in attacking any problem is like listening to a musical piece in which every note is
uniquely determined by the dominant theme. In fact, one aspect of Einstein's genius was his
astounding ability to confront the {deepest} of problems in the most {simple},
direct and {crystal-clear} manner throwing a {penetrating} light on such intricate and difficult
issues by the visionary {lucidity} of his ideas.

\bigskip

No one can deny that both the special and general theories of relativity shook the foundations of physics at
their time, giving us a new understanding of the world around us. Both theories went far beyond anything
that was well established in the world of physics.
The special theory of relativity, developed by Einstein in 1905, gave rise to a truly major revolution
in our notions of space and time. The effects
of this revolution on the known laws of physics were soon felt,
as theorists began to reformulate them within a framework compatible with special relativity.
The attempt to reformulate the theory of gravitation produced a second
major revolution in our notions of space and time; one that was even much more radical than that produced by
the special theory of relativity: Einstein's general theory
of relativity, developed in 1915 \cite{AE}. Though other scientists took part in the
discovery and development of the special theory of relativity,
the theory of general relativity was unquestionably the work of one man; Albert Einstein.
In fact, the special theory of relativity would have been discovered sooner or later
even in the absence of Einstein. The ideas in the scientific community were ripe and mature enough to give birth to
the special theory of relativity.\footnote{It should be noted that the contribution of Einstein to SR is more significant and more important than any
contribution due to other scientists (Lorentz, Poincare, \ldots.)!} {\it However,
the general theory of relativity was totally unaccepted. In fact, thanks to Einstein, it could have taken another hundred years or more for the general relativity to
be discovered! Moreover, most probably more than one person would have contributed to the theory.}

\end{mysect}

\begin{mysect}{The Special Theory of Relativity}

{\it Henceforth space by itself and time by itself are doomed to fade away into mere shadows, and
only a kind of union of the two will preserve an independent reality.} Hermann Minkowski\footnote{One of Einstein's teachers at the Institute of Technology in Zurich,
Switzerland, was the {\bf mathematician} Hermann Minkowski. Two years after the revolutionary
paper of Einstein on the nature of space and time, Minkowski made an extremely important contribution to
the theory developed by his former student. He suggested a geometric representation
for the theory of special relativity (SR) that
explained all exotic predictions of SR in a natural and elegant way. His idea was simple. Since
the Lorentz transformation upon which SR is based involves a transformation of
both space and time, one may treat time just like another dimension of space;
a forth dimension as it were. This very fruitful idea of a four dimensional space became
known as {\it Minkowski} space $M^{4}$. Addition of a fourth dimension, time, gave
a geometrical meaning to that of an {\it event}, a localization of a material
particle at a given point at a given instant. Four dimensional geometry and the
concept of four dimensional spacetime were then used to develop the laws governing the
motion of bodies. In Newtonian
mechanics, a vector $a$, for example, the momentum of a particle, is
represented by three components $a = (a_x, a_y, a_z)$. In Minkowski four
dimensional space, a vector $A$ has three space components
in addition to the time component $A = (A_t, A_x, A_y, A_z)$. The union
of space and time represented by a Minkowski space gives us a new
and deeper understanding of the laws of dynamics. For example, it turns out that the {\it four velocity} of any
particle $v = (v_t, v_x, v_y, v_z)$ representing its motion in {\it spacetime}
has norm (length) $||v|| = c$ ($c$ is the speed of light). {\it This means that all bodies move at the
speed of light in spacetime}. Accordingly, any particle at rest with respect to some
observer moves only {\it in time} at the speed of light with respect to such an observer;
a photon, on the other hand, moves only in {\it space}
at the speed of light relative to all observers. The {\it larger} the motion in space,
the {\it slower} the motion in time and vice-versa.
{\it In general, the four velocity of any body has different \lq\lq projections\rq\rq\, in
space and time as revealed to different observers.} The energy
and the momentum of a particle also acquire a totally different and much clearer
interpretation than that in the Newtonian framework. Again, they form a
single four vector; the four momentum $P = (E, p_x, p_y, p_z)$ whose norm is constant $||P|| = c^{2}$.
The energy represents the time component of $P$, whereas $(p_x, p_y, p_z)$ are the space components. Similar
to the notion of the four velocity, the four momentum of a particle may be viewed as a four dimensional vector
of {\it constant} lenght having different components with respect to different observers. {\it Both
the notions of four velocity and four momentum have the following simple and elegant geometric representation:
each may be viewed as a \lq\lq rotating\rq\rq\, four vector in Minkowski space having constant
magnitute.} Indeed, the rotation causes the projections of the four vector on the time
and space axis to vary; this is precisely the four vector veiwed from the percpective of different observers. The
constancy of the magnitude of the four vector means that the four vector transcends any particular observer.

Finally, to express the principle of relativity mathematically, we need some basic definitions.
For $u\in M^{4}$, we have $u = (u_t, u_1, u_2, u_3)$, where $u_t$ is the
time component and $u_i$, \ $i =1, 2, 3$ are the space components. The
{\it Minkowski distance} or Minkowski metric between two elements $u, v\in M^{4}$ is defined by
$\mu(u, v) := u_t.v_t - u_1. v_1 -u_2. v_2 - u_3. v_3$.
A {\it Lorentz} transformation is a {\it linear} transformation $L:M^{4}\to M^{4}$ that {\it preserves} the
Minkowski metric: $\mu(u, v) = \mu(L(u), L(v))$. A {\it Poincare} transformation is a Lorentz
transformation combined with a translation. Expressed in the geometric
language of Minkowski space, the principle of relativity can be stated as follows:
{\it Laws of nature are invariant (retain their form) under a Poincare transformation.}}

\bigskip

{\it Space is different for different observers, time is different for different observers; spacetime is the same for all observers.} John Wheeler

\bigskip

Special theory of relativity is based on the following two principles:

\begin{description}

\item [(a)] The principle of the constancy of the speed of light:
the speed of light is always the same independent of the motion of the source.

\item [(b)] The principle of relativity: All laws of physics are invariant relative to
{inertial} observers.\footnote{If one admits that there is an ultimate
velocity of signal propagation in nature, its absolute value must be the same in all inertial frames. In fact, all these frames are equivalent according
to the principle of relativity. It is impossible to suggest a physical experiment to detect the difference between them. Had the velocity of
interaction transmission been different in different inertial frames, it would have been possible to distinguish between them. This is impossible,
however, provided the principle of relativity is assumed to be universal. It follows from this that the velocity of light in vacuo must be the same in all
inertial frames of reference.}

\end{description}

Einstein chose between mechanics and electrodynamics\footnote{The actual equations that James Maxwell wrote for the electromagnetic field explain, among other things,
how light and radio waves can travel through empty space. Suppose that oscillations are set up in an electrical field. The changing (oscillating) electrical
field now acts to generate an oscillating magnetic field; located at right angles to the electrical field.
In turn, the variations in the magnetic field are the
source of oscillations in the electrical field. Accordingly, the whole thing works like a {\it feedback loop}.
The entire disturbance moves through
vacuum of space at the speed of light and is experienced as light or radio waves.}
{\it in favour} of electrodynamics. He demonstrated that
Maxwell's electrodynamics is
entirely consistent with the above two principles. Moreover, he constructed a new mechanics, {necessarily
{\it different} from Newton's} to conform
with them. He thus showed that Maxwell's theory is {\it correct} being compatible with his new theory, but Newton's
theory must be {\it modified}.

\bigskip

In his special theory of relativity, Einstein demonstrated a number of bizzare effects: masses
increase\footnote{The special theory of relativity postulates than any physical object {which has (non-zero rest) mass}
cannot move with a velocity
greater or equal to the velocity of light. A direct, though dazzling, consequence of this
postulate is that the mass of an object grows without bound
as its speed approaches that of light (the mass here is understood to be a physical quantity that
measures the resistance of the
particle to change its state of motion when acted upon by a force). This could be seen from the following reasoning:
When we apply a (constant) force to
a particle, it gains acceleration and so its speed increases. According to the above mentioned principle, its velocity cannot exceed that of light.
Therefore, the mass of the particle, increases in such a way, so that a further increase in its velocity (due to the application of the force) is
counterbalanced
by the increase in the mass of the particle. In other words, if the original force contributes to a positive increase in the velocity of the
particle, then the increase in the mass of the particle may be viewed as a hindering force - opposing the original force - that contributes
to a \lq\lq negative increase \rq\rq\, in the velocity of the particle. Both forces, the accelerating force and the de-accelerating force
(resulting from the growth of the mass of the particle) combine, so that when the velocity of particle reaches a \lq\lq critical stage\rq\rq\, \,
(acquiring a velocity comparable to that of light),  the hindering force becomes dominant and the particle starts moving with
a negative acceleration, that is, the rate of change of its velocity starts to decrease. {\it As the particle accelerates, it picks up momentum (due to
its increase of mass) and not speed!}}, clocks run slower, moving pairs of clocks get out of synchronism (time is
path dependent), rods shrink when physical systems move at speeds close to that of light and last,
but not least, that
simultaneity is frame dependent.\footnote{One of the fundamental concepts that acquired a new meaning within the framework of special relativity
is the notion of \lq\lq simultaneity\rq\rq. According
to Newtonian mechanics, it was considered absolutely obvious that events which are simultaneous in one frame are simultaneous in all frames. It
is easy to see, however,  that this statement contradicts the principle of the constancy of the speed of light. Indeed, consider two bodies $K$ and $K'$
(which we take as our frames of reference) with their corresponding clocks. Assume that $K$ moves with velocity $v$ relative to $K'$ along the
straight line joining
their centers. We place two bodies $M$ and $N$ along this line  (in the direction of motion) so that they are rigidly joint to $K'$ at equal and opposite
distances from its center. Let us consider the same process in both frames, namely the emission of a light signal from the center of body $K'$ and its
reaching
bodies $M$ and $N$. In the frame $K'$, the light signals reach $M$ and $N$ at the same instant $t'$, since both are at equal and opposite distances
from its center. In other words, {\it the two events would appear to be simultaneous in the frame $K'$.} In the frame $K$, however, $M$ moves towards
the light signal, whereas $N$ moves in the same direction as the signal. Consequently, if $t$ and $t'$ are the time taken by the light signal to reach
$M$ and $N$ respectively, then $t$ would be less than $t'$. In other words, the light signal
reaches $M$ {\bf before} it reaches $N$. {\it The two events are not simultaneous in $K$} What appears to
be simultaneous in $K'$ is no longer so in $K$.
{\it Simultaneity depends on the reference frame. Accordingly, the notion of \lq\lq now" in the
physical world is simply meaningless!}}
An immediate consequence
of this is that the same phenomenon will have different appearances relative to observers who
move at different speeds. However, Einstein
stressed the importance of an underlying unity to nature;
no matter how varried these appearances may be, {the same underlying law must hold
for all observers.} Einstein was saying that the {mathematical structure} of a physical law must {\it not} change
as we go from one observer
to the other. Laws of nature have the {\it same} form relative to {\it all} inertial observers. In fact, the essence of
the special theory of relativity may be summed in
the following statement, due to John Wheeler: {\it space} is different for {\it different} (inertial) observers, {\it time} is
different for {\it different} (inertial) observers;
{\bf spacetime} is the {\bf same} for {\bf all} (inertial) observers.

\bigskip

The second postulate of relativity theory states that \lq\lq Laws of nature remain invariant relative to all inertial reference frames\rq\rq.
Any physical law retains its structure or external form independently of the observer. Though the physical quantities
involved in one
and the same physical law may varry from one frame to another, they would always combine in a such a way
through the same mathematical
pattern. The separate components, namely, the parameters that constitute a physical law may appear different relative to
different observers, however, the
overall pattern, that is, the general mathematical form of the law remains invariant with respect to all observers.
For example, Newton's law of motion
remains valid as a law that describes the motion of a particle under the action of a force. However, each physical
quantity appearing in
Newton's law acquires a drastically new meaning in the framework of special relativity theory.
The formulation of such a law combines the following
concepts: the force acting on the particle, the momentum of the particle (the product its mass and velocity) and
its rate of change.
Each of these quantities are not absolute, but change from one reference frame to the other
(for example, the mass of a moving body is always greater
than its rest mass). However, and this is the important point, all observers, no matter what position
they stand relative to the particle, would describe the
change in motion of the particle as the result of the action of a force acting on the particle. Though, the force
differs relative to each observer, {\it the mathematical formula that expresses such a law
acquires an absolute status in so far as it transcends the relative position of the
subject.}  Relativity theory, contrary to what is believed, is an {\it objective} theory per-exellence
in the sense so far discussed. Such an objective
character of the theory may be better apprehended if one borrows Einstein's own words to describe it as the
\lq\lq the theory of invariants.\rq\rq. Indeed,
it is true that the special relativity theory does introduce the position of the observer, making it play a
fundamental role in its very formulation. {\it However, it also succeeds, inspite
of emphasizing the role played by the subject, to give a clear, unambigous meaning of an
objective physical world that transcends the very existence of
the subject. In fact, the second postulate of the theory can be regarded as
a definition of the meaning of objectivity in the classical sense.}

\end{mysect}

\begin{mysect}{The General Theory of Relativity}

Some of the material discussed here may be found in \cite{De}.

\bigskip

{\it In 1919, Einstein's nine-year old son Edward asked him: \lq\lq Daddy, why are you so famous?" Einstein laughed
and then explained quite seriosly: \lq\lq You see, son, when a blind bug crawls along the surface of a
sphere it doesn't notice that its path is curved. I was fortunate enough to notice this".}

\bigskip

{\it I hold in fact (1) That small portions of space are in fact of a nature analogous to little hills on a surface which is on the average flat; namely, that
the ordinary laws of geometry are not valid in them (2) That this property of being curved or distorted is continually being passed on from one portion
of space to another in the manner of  a wave (3) That this variation of the curvature of space is what really happens in that phenomenon which we call
the motion of matter.} {William Clifford} as quoted in \cite{FDP}.

\bigskip

{\it It is a matter of fact that Leibniz applies his principle successfully to the problem of motion and that he arrived at relativity of motion on logical
grounds..... The famous correspondence between Leibniz and Clarke... reads as though Leibniz had taken the arguments from expositions of
Einstein's theory.} Hans Reichenbach, \lq\lq The philosophical significance of relativity\rq\rq.

\bigskip

There is no doubt that the intellectual leap accomplished by Einstein to move from the special to
the general theory of relativity is one of the {\it greatest} in the history of human thought.
There are {\bf five} principles which, explicitely or implicitely, guided Einstein in his attempt to generalize
his special theory of relativity to a more general theory that
encompasses gravity.\footnote{The {\it logical core} of the special theory of relativity
(the relativity of space and time) is simply in {\it direct conflict}
with the {\it logical core} of Newton's theory of gravitation (based on the assumption that space  and time are absolute).
It is strange that no one other than Einstein struggled to modify Newton's theory to make it compatible with his
special theory of relativity. This actually
took Einstein eight years of hard work!}

\bigskip

We first recall these five principles, then we elaborate as we go along.

\begin{description}
\item [(1)] Mach's principle

\item [(2)] Principle of equivalence

\item [(3)] Principle of covariance

\item [(4)] Principle of minimal coupling

\item [(5)] Correspondence principle

\end{description}

\bigskip

The status of these principles has been the source of much controversy. For example, the principle of covariance
(laws of nature should be
expressed in the language of tensors) is considered by some authors to be empty,
whereas others claim that it is possible to derive general
relativity more or less from this principle! The status of this principle, we believe, lies somewhere in between these
two extreme views. One can say that tensors are
one possible mathematicl formalism (among other approaches)\footnote{One such
approach is due to Istvan Nemeti and Hajnal Andreka, the Hungerian Professors at the mathematical
institute of Hungarian Academy of Science, together
with their student Judit Madarasz. They have succeeded to formulate the special theory of
relativity in an axiomatic method based on
a transparent relatively simple set of axioms. Towards general relativity, they played with
models where observers are allowed to accelerate. At the limit,
using ultra products or non-standered analysis, they obtained two notions which are two sides of the
same coin. One is {\bf gravity}, or rather
the logistic formulation of a four dimensional manifold ({\bf curved} spacetime) which is basically a
geometry in the very
broad sense of the word. The other is {\bf acceleration}, viewed as a delicate patching of instantaneous inertial frames.
The {\bf duality} between gravity
and acceleration, in short, the {\bf equivalence principle}, is thus formulated as a typical {\bf adjoint situation}
that pops up in different parts of
mathematics \cite{PHD}. This approach to general relativity uses the machinary of algebraic logic in formulating
Einstein's field equations \,\cite{PR} {\it avoiding the use of tensors!} The goal of the project is to prove strong theorems
of the special and the general theory of
relativity from a small number of easily understandable and convincing axioms. In doing so, the authors try
to eliminate all tacit assumptions from relativity and replace them
with explicit and crystal clear axioms in the spirit initiated by Tarski in his first order axiomatization of
geometry \cite{Tar}.} formulating
Einstein's general theory of relativity. There is a fairly general
agreement, however, that among the five principles, the equivalence principle is the {\bf key} principle.
\footnote{The idea of abolishing gravitational
effects by the process of free fall was considered
by Einstein as the \,\lq\lq happiest thought of my life \rq\rq \cite{AE}. Einstein was aware that such a simple idea
would be a breakthrough in generalizing his special theory of relativity to a more general theory of gravity: {\it gravity
cannot be abolished globaly, but only locally!}.}
It has a special status. Indeed,
the mere fact that Einstein built the theory on this principle justifies its importance.
One source of the different views concerning the other four principles
arises from the fact that these principles are more of a {\bf philosophical} nature.
Accordingly, they are not on an equal footing with the
equivalence principle in so far as they invoke controversy.\footnote{The geometric langauge of the
general theory of relativity is
$4$-dimensional Riemannian geometry. Though it
is the best known theory for studying gravitational interaction, so far, it suffers from some problems.
Examples of these problems are: the
horizon problem, the initial singularity, the flatness of the rotation curve of spiral galaxies \cite{SG}, the
Pioneer 10, 11-anamoly \cite{JJJ} and the interpretation of supernovae type-Ia observation\,\cite{JJ}. Some of
these problems are old (for example, the initial singularity and the horizon problems), while
others have been discovered in the last ten years or so. In formulating general relativity (and possible
generalizations), some authors prefer to use more general geometric structures
(cf. \cite{G.S.A}, \cite{Gh}, \cite{WNA}, \cite{b}, \cite{MIWz},
\cite{AMR}, \cite{EAP}). The author of this paper, together with Professor Nabil L. Youssef and
Professor Mamdouh I. Wanas, constructed a unified field theory ($UFT$) unifying gravity and electromagnetism
and possibly other interactions in a much richer context
than Riemannian geometry (\cite{WNAL}, \cite{linear}). Unlike the classical general theory of relativity,
which is formulated on the base manifold $M$ (spacetime manifold of dimension $4$), the $UFT$ is constructed
in the tangent bundle $TM$ of $M$ (a manifold of dimension $2\times 4 = 8$). These \lq\lq extra
degrees of freedom\rq\rq \,make the $UFT$ \lq\lq wide enough\rq\rq\,\, to describe
electromagnetic interactions on a {\bf geometric} basis giving matter a {\bf geometric}
origin. It is still an open question, though, whether the $UFT$ is
capable of achieving a {\it geometric} description of other {\it micro}-phenomena.}

\subsection{Mach's Principle}

The writings of Ernst Mach, the great nineteenth-century physicist and philosopher, had deeply
influenced the young Albert.
Mach not only made significant contributions to the foundations of physics, but his writings also deeply
influenced the logical
postivists.\footnote{A school of thought that played an important role in the
development of science. They believed that the reason for the lack of progress in the
domain of philosophy is that most philosophers use ambiguous, abstruse and vague concepts in tackling
fundamental questions. As such, they regarded most such doctrines of thought as having
a negative influence on scientific knowledge hindering its progress. {\it Considering mathematics and
physics as the only exact sciences, the logical postivists struggled to eliminate \lq\lq metaphysical" ideas
from the world of science and philosophy.} They claimed that such ideas were not well defined sometimes even
paradoxical and thus may lead to unresolved contradictions.
{\it Their moto was: define your terms before using them!} In fact, Einstein in his early days was
largely influenced by such a school. His paper on the special theory
of relativity was written to a large extent in the spirit of the above motto. Years later, though,
Einstein regarded logical postivism as somewhat rigid and unflexible. He realized that
metaphysical ideas could be useful and effective in the development of scienctific thought. In formulating
new ideas, one needs a margin of ambiguity. Ideas should not be strictly defined before hand,
they are not \lq\lq born\rq\rq in the best possible form, but acquire articulation and exactness as they
evolve. They are dynamic with no well defined boundaries.}
He was the kind of philosopher who was always asking troublesome questions. Mach had been
particulary puzzled by the meaning of certain properties like linear and angular momentum
{\it in an otherwise empty space}.
In fact, Einstein specifically referred to this question of Mach\footnote{Sad to relate, Mach in his old age,
after his insights had been incorporated by Einstein into a successful theory,
refused to accept relativity.}  in the first pages of the great paper \lq\lq
The Foundations of the General Theory
of Relativity \rq\rq. What, Mach had asked, would it mean to say that a planet is rotating in empty space; that is, if
there was
nothing else in the universe against
which this rotation can be measured? For example, if the sky were totally empty, how would
we ever know that the earth spins on its
axis?\footnote{Roger Penrose asked the same question while constructing his
twister theory \cite{FDP}. Quantum theory gives a meaning to the
spin of the electron. It claims that an electron can spin in one of two alternative directions: \lq\lq up\rq\rq \, or \lq\lq
down\rq\rq. \, But what meaning would these alternatives
have when the universe is totally empty? What would the difference be between a spin up and a spin down?
We should expect such differences
to manifest themselves only when a number of other reference points are present. If the distinction
between a spin up and a spin down is to
have meaning within a quantum theory set in empty space, this seems to imply that rather than living in
some sort of a general background space,
electrons actually create their own spacetime. Each electron would therefore have an associated \lq\lq primitive
space\rq\rq\, - possibly at this stage nothing
like our own spacetime at all. However, as large numbers of electrons combine, Penrose conjectured,
it is possible that such individual \lq\lq protospaces\rq\rq\, \, to
fuse together, giving rise to a collective space; a sort of shared spatial relationship which may begin to resemble
our own space. Penrose was
actually using Mach's ideas now in the broader context of a quantum theory of gravity: the properties of
space(time) are {\it both a reflection and a result} of the properties of the objects living in space(time).}
It is possible to answer that the earth
bulges at the equator because of the effect of centrifugal force. Because we can observe this bulge, it implies
that the earth must indeed be spinning. But what exactly is this centrifugal force, and
how and why does it arise? If we choose a set of axis that
rotates at exactly the same speed as the planet, then everything would appear stationary in this
otherwise empty space. Where does this force
have its origin? What does it mean for there to be a direction of spin when nothing else is present?

\bigskip

In short, Mach's principle means that there is no meaning to the concept of {\bf absolute} motion, but only
that of {\bf relative} motion.\footnote{Leibniz,
among Newton's critics, saw most deeply why Newton's conception of
absolute space and time could not ultimately succeed. The argument Leibniz
makes for his relational point of view is one of the most important in the whole history of
philosophizing about nature. We cannot do better than reproduce
his words: \lq\lq I am granted this important principle that nothing happens without a sufficient reason why
it should be thus rather
than otherwise...I say
then that if space were an absolute being, there would happen something for which it would be impossible
that there should be a
sufficient reason, and
this is contrary to our axiom. This is how I prove it. If we suppose that space is something in itself,
other than the order of bodies among themselves, it
is impossible that there should be a reason why God, preserving the same position for bodies among themselves,
should have arranged bodies in
space thus, and not otherwise, and why everything was not put the other way round (for instance) by changing
east and west.} \,In a populated universe,
it is the interaction between the matter in the universe - other and above the gravitational interaction -
which is the source of inertial effects. In our universe,
the bulk of the matter resides in what is called the fixed stars. An inertial frame is a
frame in some previleged state of motion relative to the average motion of the
fixed stars. {\it More specifically, each and every body is {coupled} to the
whole universe through its interaction with all other bodies.}

\bigskip

Leibniz taught us to reject any reference to {\it a-priori} and immutable structure, such as Newton's absolute space
and time. But
he did not tell us what to replace them with. Mach did, for he showed us that every use of such an absolute entity
hides an
implicit reference to something real and tangible that has so far been left out of the picture.
What we feel pushing against us
when we accelerate cannot be absolute space, for there is no such thing. It must somehow be the whole
of the matter of the universe.\,Einstein
took a third step in the transformation from an absolute to a relational conception of space and time.
In this step, the absolute
elements, identified by Mach as the distance galaxies, are tied into an interwoven, dynamical cosmos. {\it The final
result is that the
geometry of space and time - which was for Newton absolute and eternal - became dynamical, contingent and lawful.}

\bigskip

Let us now return to the relational viewpoint of space and time. Space and time are merely bookkeeping
devices for conventiently summarizing
relationships between objects and events in the universe. The location of an object in space has
meaning only in comparison with another. {\it Space
and time are the vocabulary of these relations and nothing more}. The above ideas may
be succintly summarized as follows:

\bigskip

$M_1$: Matter distribution determines the geometry. By the geometry of the universe is meant the
privileged paths which particles and light rays travel.

\bigskip

$M_2$: If there is no matter there is no geometry.

\bigskip

$M_3$: A body in an otherwise empty universe should possess no inertial properties.

%\newpage

\subsection{The Equivalence Principle}

The equivalence principle may be summarized in the following four principles:

\bigskip

$P_1$: The motion of a gravitational test particle \footnote{A particle which
experiences a gravitational field without altering the field.} in a gravitational field
is independent of its mass and composition.

\bigskip

This is known as the {\bf strong} form of the equivalence principle. In Newtonian theory, it is an observational result;
a coincidence. It is possible and
compatible with Newton's theory of gravity that if we look closer, with an accuracy greater than
$1$ in $10^{12}$, different bodies would undergo
different accelerations when placed in a gravitational field. {\it In general relativity, however, it forms the essential
{logical core} of the theory; if it falls then so
does the theory.}

\bigskip

$P_2$: The gravitational field is coupled to everything.

\bigskip

This is known as the {\bf weak} form of the equivalence principle. It makes explicit the assumption that matter
both response to, and is a source of gravity.
In other words, {\it no body is shielded from a gravitational field}. However, it is possible to remove
gravitational effects (and hence regain special relativity)
by considering a {\bf local} inertial frame; a frame in a state of {\bf free}-fall.\footnote{A reference frame is
inertial in a certain region of space and time when throughout that region of spacetime
and within some specific accuracy, every test particle originally at rest with respect to that frame
remains at rest, and every
free test particle initially in motion
with respect to that frame continues its motion without change in speed and direction.} This
naturally leads to the following:

\bigskip

$P_3$: There is no local experiment that distinguishes a non-rotating free-fall in a gravitational field from
uniform motion in space in the absence of
a gravitational field.

\bigskip

Einstein noticed another {\it coincidence} in Newtonian theory. All {\it inertial forces} are proportional to the mass
of the body experiencing them. The gravitational
force has the same property. This led Einstein to conclude that gravitational (inertial) effects are inertial
(gravitational) effects. \footnote{This means that
gravity and acceleration are actually {\bf indistinguishable}. The distinction between who is accelerating
and who is not may be thought of as part of the
intrinsic structure of space and time. For Newton, it was absolute. Mach and Einstein made it dynamical;
the distinction can be made differently at
different places and at different times.} Put differently,
{\it gravitation is an effect that arises from not using an inertial frame.} Consequently, we have

\bigskip

$P_4$: A frame linearly accelerated relative to an inertial frame in special relativity is {\bf locally}
identical to a frame at rest in a gravitational field.

\bigskip

The last two versions of the equivalence principle can be {\bf vividly} clarified by considering
Einstein's famous {\bf lift} experiment:\footnote{Einstein always loved simple examples, pictures that one might consider with suspicion on
account of their apparent naivete; for example, of trains,
boxes, rooms and lifts. One aspect of Einstein's genius was his astounding ability to express deep and
subtle ideas through the invention of
\lq\lq simple\rq\rq and sometimes apparently naive thought experiments. Einstein devised a
thought experiment that illuminated one of the most important predictions of the
special theory of relativity; namely, the {\bf equivalence} of mass and energy.
(Einstein demonstrated that the energy $E$ of a particle and its
mass $m$ are not independent quantities but are related by the equation $E = mc^{2}$, where $c$ is the speed
of light. This is probably the most famous equation in
modern physics!). Consider two identical particles (i.e. having the same rest mass)\,
moving
relative to an observer in a straight line towards each other with equal opposite velocities, say $v$ and $-v$.
As the mass depends on
the (square of the) velocity, both particles will have the same mass $m$ relative to the observer, which is,
according to the predictions of the
special theory of relativity, greater than their rest mass (if $m_o$ is the rest mass of each particle,
then $m = m_o/(1 - (v/c)^{2})^{1/2}$, where
$c$ is the speed of light). Since both particles move with the same velocity in two opposite directions,
the total momentum
of the system vanishes (the momentum is a vector quantity which
depends on the direction of motion. Hence, if the momentum of the first particle is $mv$,
then the momentum of the second particle, owing to the fact that
it is moving in the opposite direction, is given by  $-mv$. The total momentum measured by the observer is
thus found to be $mv - mv = 0$).
The law of {\it conservation} of momentum tells us that the total momentum of a closed system, i.e. any system
of particles in which no external forces act,
does not change with time. This implies that the total momentum of the system under consideration
before, after and at the instant of collision remains identically zero. In particular, at the instant of collision, the two
particles appear to the observer to
be {\it momentarily at rest}, and so he must be observing their {\it rest mass} $m_o$. Accordingly, the total mass
of the system before (and after) collision appears to
be {\it greater} than its mass at the instant of collision. This, however, violates the principle of
{\it conservation} of mass. In fact, the difference in mass,
according to the principle of equivalence of mass and energy, is lost or transformed into the energy of
deformation  and the internal frictional forces
resulting from the collision of the two particles. A glaring example of a (very simple!) physical situation in
which mass is transformed into energy. It
should be noted, though, that the previous arguments are not a proof of the equivalence of mass and energy,
but merely a manifestation of such a principle.

\par

Another brilliant thought experiment was introduced by Einstein in his passage from the
special to the general theory of relativity.
Einstein deviced a thought experiment
showing that the {\bf curving} of space(time) is in fact demanded when (constant) {\bf accelerated} motion is {\bf combined} with
(the predictions of) the special theory of relativity (SR + ACC$\implies$ CUR)\,\cite{PR}. Take the case
of a rotating disc such as the platform of a merry - go - round. Contemplation of the seemingly
simple mechanics of such a disc actually provides
deep conceptual insight. {\it It is a stepping stone for the transition from the special to the
general theory of relativity}. In fact, this thought experiment was a turning point in Einstein's success to
generalize his special theory of relativity to a theory of gravitation. We are assured that the
special theory of relativity is valid relative to an inertial frame such as the
non-rotating frame outside the merry go round, which we denote by $R$.
Relative to $R$, and
using meter sticks {\it at rest} in $R$, the diameter $D$ and the circumference $C$ of the merry-go-round
have the Euclidean ratio $C/D = \pi$.
In the {\it rotating} frame
$R'$, however, points on the rotating platform are {\it accelerating}. But if the platform is large enough,
then a point on the rotating circumference $C'$
moves {almost in a straight line for a short period.} We can therefore have a meter stick that
moves exactly in a straight line (with constant velocity) move along
with this point in such a way that it has for that instant exactly the same velocity as the point on $C'$.
In this way, we can calibrate meter sticks at rest on
$C'$ with meter sticks moving uniformly in $R$. This permits us to transfer our standered of
lenghts from the inertial frame $R$ to the
non-inertial frame $R'$.  According to the predictions of the special theory of relativity,
the uniformly moving meter stick will be contracted by a factor $1/\gamma$, where
$\gamma = 1/(1 - (v/c)^{2})^{1/2}$. Therefore the
standard at rest on $C'$, which is a copy of it, will be also contracted relative to the one at rest in $R$.
This means that the observer in $R'$ will measure
along his circumference $C'$ with a {\it shorter} meter stick than the observer in $R$.
Consequently, the observer in $R'$ finds the circumference
$C'$ to be {\it larger} than $C$ by a facter $\gamma$. On the other hand,
the radial distance on the platform {\it will not be contracted} because it is
{\it perpendicular to the direction of motion.} The diameter $D'$ as measured by the observer in $R'$ will
thus be the same as the diameter $D$ mearured
in $R$. Conclusion: the ratio of circumference to diameter in the rotating frame
$R'$ is $C'/D' = \gamma C/D = \gamma \pi$, which is {\it larger} than
plane Euclidean geometry permits. This led Einstein to propose the following  {\bf amazing} idea:
{\it the {\bf curving} of {\bf space} is the (explanation for the) violation of ordinary Euclidean geometry}.
The above argument concerning {\it length} measurments can easily be repeated for {\it time} measurments.
One replaces the meter stick at rest on
$C'$ by a clock, and one compares that clock with one that is uniformly moving relative to $R$.
One finds again, using the same reasoning as above,
that the clock at rest in $R'$ goes {\it slower}
than the clock at rest in $R$ (time dilation). Moreover, the {\it larger} the platform, the {\it slower}
will be the clock on its circumference.
Again {\bf time} is {\bf wrapped} : {\it its rate of passage differs from one location to another.}}

\bigskip

Case 1: The lift is placed in a rocket ship in a part of the universe far removed from gravitational influence.
The rocket is accelerated forward
with constant acceleration $g$ relative to an inertial frame. The observer in the lift
releases a body from rest and
(neglecting the influence of the lift) sees it fall to the floor with acceleration $g$.

\bigskip

Case 2: The rocket engine is switched off so that the lift undergoes uniform motion relative to the inertial observer.
A released body is found to remain at rest relative to the observer.

\bigskip

Case 3: The lift is next placed on the surface of the earth, whose rotational and orbital motion are ignored. A released body is
found to fall to the floor with acceleration $g$.

\bigskip

Case 4: Finally, the lift is placed in an evacuated lift shaft and allowed to fall freely towards the center of the earth.
A released body is found to remain at rest relative to the observer.

\bigskip

It is clear that $P_3$ implies that cases $1$ and $3$ are indistinguishable, whereas $P_4$ implies that cases $2$ and $4$ are indistinguishable.

%\bigskip

\bigskip

To sum up, the equivalence principle asserts that it is impossible to tell whether one is in a room which is
freely falling in a gravitational field or
in a room moving uniformly in deep space. It also asserts that a room which is accelerating steadily in deep space,
with the same acceleration as falling bodies at the
surface of the earth have, is indistinguishable from a room sitting on the surface of the earth.\footnote{Many
beautiful effects follow from this
simple principle, such as the bending of light and the slowing down of clocks in a gravitational field \cite{a}.}

%\newpage

\subsection{The Principle of Covariance}

The principle of relativity, which lies at the heart of special relativity, tells us that {\it al physical laws
must be the same regardless of the
constant-velocity relative motion that individual observers might experience}. This is a {\it symmetry} principle, because it means that nature treats
all such observers identically - {\it symmetrically.} Through the {\it equivalence principle} - the corner stone of the general theory
of relativity - Einstein significantly extended this symmetry principle by showing that {\it the laws of physics are actually identical for all observers
regardless of their state of motion: an accelerated observer is perfectly justified in declaring himself to be at rest and claiming that the force he
feels is due to a gravitational field.} {\it Once gravity is {included} in the framework, all possible observational vantage points are on a completely
equal footing.}\footnote{Just as the symmetry between all possible observational vantage points in general relativity requires the existence of
the gravitational force, developments relying on work of Hermann Weyl in the $1920$'s and Chen-Ning Yang and Robert Mills in the $1950$'s showed
that gauge symmetries {\it require} the existence of yet other force fields. Certain kinds of force fields, according to Yang and Mills, will provide
perfect compensation for \lq\lq shifts in force charges \rq\rq, thereby keeping the physical interactions between the particles completely unchanged. For
the case of the gauge symmetry associated with shifting quark-colour charges, the required force is none other than the strong force. A similar discussion
applies to the week and electromagnetic forces, showing that their existence, too, is bound up with yet other gauge symmetries - the so called weak and
electromagnetic gauge symmetries. And hence, all four forces are directly
asociated with principles of symmetry. \cite{Brian}}  Einstein thus proposed the following as the
logical completeness of the principle of special relativity:

\bigskip

Laws of nature should have the {\bf same} form relative to {\bf all} observers.

\bigskip

Observers are intimately tied up with their reference systems. If an observer can discover a physical law, than any other observer (no matter
what kind of motion he is experiencing) will discover the same law. In other words, any coordinate system should do. The situation is different in
special relativity.\footnote{Technically speaking, because the connection is integrable and the metric is flat, there exists a canonical coordinate system, namely
Minkowski coordinates. In a curved spacetime, that is, a manifold with a non flat metric, there is no canonical coordinate coordinate system}
If spacetime is curved, then there are no frames that qualify as being inertial everywhere. Free falling frames of reference are coordinate systems
whose axis are straight only in the vicinity of a point locally. If extended beyond this region, they have no properties that would
distinguish them from other curvilinear coordinate systems. In a curved spacetime manifold any coordinate system is as suitable as
any other.\footnote{This statement should be treated with caution. In many applications there would be a preferred coordinate system. Many
problems possess symmetries. In this case, it is advisable to adapt the coordinate system to the underlying symmetry.} {\it It is not so much that
any coordinate system should work, but rather that the theory should be invariant under a coordinate transformation.} This is expressed mathematically
as follows:

\bigskip

The equations of physics should be expressed in the langauge of tensors.\footnote{Any tensorial equation
remains invariant (retain its form) under a change of coordinate.}

\bigskip
%\newpage

\subsection {Principle  of Minimal Gravitational Coupling}

The minimal gravitational coupling is a {\bf simplicity} principle. We should not add
unnecessary terms in the transition from the special to the
general theory of relativity. For example, in the
context of the special theory of relativity, the energy momentum tensor satisfies the conservation law
$$T^{ab}\,_{, b} = 0,$$
where \lq\lq , \rq\rq denotes partial differentiation.\footnote{Application of equivalence principle
gives same equation in a local Lorentz frame.
Since the connection coefficients vanish at the origin of any local Lorentz frame, the above equation may
be written {\it in any reference frame}
in the form $T^{ab}\,_{;b} = 0$, where \lq\lq ; \rq\rq denotes covariant differentation
(unlike partial differentation, the covarient derivative of a tensor
is a tensor). Thus we arrive at the following remarkable result:
The laws of physics written in component form change in passage from {\it flat} spacetime to
{\it curved} spacetime by a mere {\it replacement of all commas by semi-colons}. No change
in physics; change due to a switch in reference
frames from Lorentz to non-Lorentz! \cite{a}.} The simplest generalization of the above
equation in {\bf curved} spacetime is the following tensorial equation
$$T^{ab}\,_{; b} = 0, $$
where \lq\lq ; \rq\rq denotes {\bf covariant} differentation.\footnote{We face here a
technical problem. In what order should the derivatives be written
when applying the commas goes to semi-colon rule? Interchanging derivatives makes no difference in flat spacetime.
In curved spacetime, however,
{\it it produces terms that couple to curvature!} This is a non-trivial issue \cite{a}.}

\bigskip

Adding new terms containing the curvature tensor to the last equation {\it will not alter the form of
the first equation}. For example, we can write
$$T^{ab}\,_{; b} + g^{be}R^{a}_{bcd}T^{cd}\,_{; e} = 0$$
This is because the
Riemann curvature tensor $R^{a}_{bcd}$ - the mathematical object describing gravity,
vanishes in the context of special relativity. We can thus add terms explicitely containing the
curvature tensor when passing from the s   pecial to the general without violating the
form of the equations obtained in the special relativity.
Accordingly, Einstein implicitely applied the following principle:

\bigskip

Principle of minimal gravitational coupling: No terms explicitely containing the curvature tensor should be added in making the transition
from the special to the general theory of relativity.

%\newpage

\subsection{The Correspondence Principle}

Any new theory must be consistent with any established earlier theory. Thus general relativity must agree with both special relativity (in the
absence of gravitation) and Newton's theory of gravity (in the limit of weak gravitational field and low velocities). In fact, general relativity has
as distinct approximate theories:

\begin{description}

\item [(a)] Special relativity: General relativity has two distinct kinds of correspondence with special relativity. The first is the
limit of vanishing curvature (flat spacetime). In this case, one can introduce a global inertial frame and recover completely the
special theory of relativity. The second is local rather than global. In a local inertial frame, all the laws of physics takeon their special
relativistic form (the equivalence principle).

\item [(b)] Newtonian theory: In the limit of weak gravitational fields, small velocities (compared to the speed of light), and small pressures and density
of matter, general relativity reduces to the Newtonian theory of gravity.

\item [(c)] Post Newtonian theory:  When first order relativistic corrections are added to the Newtonian theory, that is when Newtonian theory is
nearly valid, one deals with Post Newtonian theory of gravity.

\item [(d)] Linearized theory: In the limit of weak gravitational fields, but possibly large velocities and pressures, general relativity reduces to the
linearized theory of gravity.

\end{description}

\subsection{Black Holes}

We now discuss one of the most important predictions of Einstein's general theory of relativity. The most
{\bf dramatic} prediction of Einstein's
theory are {\bf black holes}. A detailed and excellent treatment of the geometry of black holes may
be found in \cite{a}. See also \cite{Haw}.

\bigskip

The misleading name (invented by John Wheeler) does not refer to holes, but to regions in space where the gravitational pull
is extremely strong, stronger than a certain {\it critical} value. The critical value can be characterized as follows: One of the important predictions of
the general theory of relativity is that light rays are bent by light.\footnote{The bending predicted by general relativity is greater than that predicted by
Newton's gravitation.} Actually, the bending depends not only on the mass $m$ but also on the
distance $d$ from the center of that mass at which the light ray passes. The smaller the distance, the larger is the bending. In fact, the bending depends on
the ratio $m/d$. As stars burn out, as they deplete energy so that they cannot radiate light any more, the gravitational pull of its matter
collapses the star to a very compact \linebreak \lq\lq sphere\rq\rq of extremely high density. As a result, the ration $m/d$ increases tremendously: A collapsed star
can bend light rays a great deal more strongly because the rays pass by it at a much closer range. If a \lq\lq dead\rq\rq \,star collapses to a such
a small size that the light rays reaching its surface are bent into the interior of the star so that they cannot escape again, it is called a black hole. If light
rays get caught so does everything else (travelling at lower speeds). Now a black hole does not send out radiation of its own; therefore if it
absorbs all radiation reaching it and reflects none, it is necessarily invisible. The region of space appears entirely black.\footnote{{\it This is true {\bf only} in
the context of the classical general theory of relativity}. Stephen Hawking, the brilliant Cambridge physicist, argued that close to a black hole, the extreme degree
of curvature of spacetime actually creates elementary particles. {\it In more detail, taking the quantum theory into consideration, Hawking
showed that black holes actually {radiate} light, so that they
are not too black after all}!  The idea is simple though the calculations are quite tedious \cite{Brian}. The uncertainty principle, the corner-stone of the quantum
theory, tells us that space is a teeming rolling frenzy of virtual particles, momentary errupting into existence and subsequently annihilating one another.
This jittery quantum behavour also occurs on the event horizon of a black hole. Hawking realized, however, that the gravitational might of the black hole
can inject energy into a pair of virtual photons. This process tears them just far enough so that one gets sucked into the hole. With its partner having
disappeared into the black hole, the other photon of the pair no longer has a partner with which to annihilate. Instead, the remaining photon gets an
energy boost from the gravitational force of the black hole and, as its partner falls inward, it gets shot outward, away from the black hole. The combined
effect of such a process, happenning over and over again around the horizon of the black hole,
appears as a steady stream of outgoing radiation. Black holes have temperature; hence have {\it entropy}. A {\bf partial}
unification of relativity and quantum theories accomplished by Hawking gave rise to a radically new and
totally unexpected description of black  holes: "black" holes can actually "glow"! {\it In fact, Hawking's discovery
says that the gravitational laws of black holes are
nothing but a rewriting of the laws of thermodynamics
in an extremely exotic gravitational context! This was Hawking's {\bf bombshell} in 1974}. (Incidently, the above
phenomenon is an indication that the uncertainty principle (UP) is, in a sense, {\it not} a \lq\lq limitative" result after all.
GR without the UP predicts that black holes are completely invisible. On the other hand, GR with UP (as studied by Hawking) predicts
that black holes can glow: a combination of general relativity with (the statistical nature of) the quantum theory (applied to
a \lq\lq black" hole) gives the following amazing and astounding result: $GR + UP\implies$ {\it
Black holes have entropy. They can glow).}}

\bigskip

We now consider the collapse of a {\bf spherically} symmetric non-rotating star until its surface reaches its
{\it Schwarzchild} raduis.\footnote{The Schwarzchild
raduis of the earth is $1$ cm and that of the sun is $3$ km.} As long as the star remains spherically symmetric, its external field is given by
the Schwarzchild vacuum solution.\footnote{Technically, this is known as Birkhoff's theorem: Let the geometry of a given region of spacetime be
spherically symmetric and be a solution to Einstein field equations in vacuum. Then that geometry is necessarily a piece of the Schwarzchild
geometry \cite{a}.} If signals are sent out an observer on the surface of the star at {\it regular} intervals according to his clock, then as the surface of the star reaches
the Schwarzchild raduis, a {\it distance} observer will receive these signals with {\it an ever-increasing time gap between them}. In particular, the signal
sent off at $r = 2m$ will never escape from $r = 2m$, and all successive signals will ultimately be dragged back to the singularity at the center. No matter
how long the distant observer waits, it will be only possible to see the surface of the star as it was just before it plunged through the Schwarzchild
raduis. It follows that there is a great difference in the description of motion of both observers. From the point of view of the {\it falling} observer, it takes
only a {\it finite} time to reach $r = 2m$ or even $r = 0$. However, relative to the {\it far away} observer, safely outside the event horizon, it takes an
{\it infinite} amount of time for the falling observer
to reach $r = 2m$. The whole range of $r$ values, from $r = 2m$ to $r = 0$ is perfectly good physics, and physics that the falling observer is going to see
and explore, but physics that the outside observer never will see and never can see. This phenomena is a clear
example of what may be called
\lq\lq infinite time dilation\rq\rq. Note that this occurs in the context of stationary (static) black holes.
It is an inevitable consequence of Einstein's field equations!\footnote{The phenomenon of infinite time dilation
also shows that {\bf different}
observers in the context of GR may {\bf not}
see the same set of events. Indeed, the falling(in) observer passes through the horizon of the black hole
reaching the singularity at the center of the black hole in a {\it finite} time according to his clock (realistically,
of course, the falling observer will be torn to pieces (crushed!) as he gets closer and closer to the horizon).
On the other hand, the far-away observer will never \lq\lq see" {\it beyond} the horizon of the black hole.
All events interior to the horizon simply do {\it not} belong to his world view. From his
\lq\lq percpective", the far-away observer will
\lq\lq think" that the falling observer is coming to a {\bf halt} as he approaches the horizon of the black hole and
will never see him crossing the horizon.}

\bigskip

Andreka and Nemeti showed that if one takes into
account the laws of the general theory of relativity, more specifically, the infinite time dilation in strong
graviational fields of (rotating) black holes,
then one can imagine \lq\lq thought experiments\rq\rq\, in which Church thesis is no longer valid!
In these \lq\lq thought experiments\rq\rq, one can
{\it compute non computable functions (in the old sense) and one can prove that $ZFC$ is consistent}!
So was David Hilbert, right after all?
\footnote{Five days before Einstein presented his field equations in its final form, Hilbert animated
by Einstein's earlier work, independently
discovered how to obtain them from an action principle\cite{a}.} Ironically, it
was Kurt Godel who started considering such thought experiments involving rotating black holes with
strong gravitational fields.\footnote{In the 1949
volume celebrating Einstein's seventieth birthday, Godel presented work that sparked research aimed at
finding exact solutions to
Einstein's field equations that were more complex than any previously known.
In particular, Godel's solution allowed {\it closed time like world lines},
that is, it allowed time to be {\it cyclic!}. The solution was genuinly puzzling and
raised very deep questions concerning the nature of time. In fact,
Einstein regarded it as too absurd and thought that it should be excluded on physical grounds.}
Obviously, Godel was not concerned with violating his own incompleteness results! \footnote{Church thesis is
the assumption that the class of recursive functions coincide with the class of computable functions.
Godel proved that any recursive axiomatization (via Godel numbering)
of arithimetic is not complete. This is known as his first incompleteness theorem.
His second incompleteness theorem is even more profound, with devastating philosophical implications.
It states that any strong enough system (like Peano arithmetic $ZFC$) cannot prove its own consistency,
unless it is inconsistent in which
case it can prove anything.
So there is always the possibilituy that $ZFC$ turns out to be inconsistent, though highly unlikely.
Returning to recursive and computable functions,
Turing gave an equivalent definition of computable functions as those functions that a Turing machine can compute.
Other rigorously defined clases of functions, that turn out equivalent to
recursive function are given by the Russian  mathematician Markov, and Church via his invention of the so called
lambda calculas. In fact the Lambda calculas was an outcome of Church's attempts to vilate Godel's first incompleteness theorem,
via a system that Kleene proved inconsistent. Markov's procedure is now termed Markov's algorithms
highlighting the algorithmic character of
computing recursive  functions. Today a recursive function is one that a computer can determine.
A historical comment:
When Godel proved his first incompleteness result he dealt only with the now called primitive recursive functions,
which he attributed the name recursive to them. This
is a class that is strictly smaller than that of the now called recursive functions, which Church
introduced as an attempt to violate Godels first incompleteness theorem thinking that
it does not  apply to the wider class of the now called general recursive functions.
Godel however showed that his reasoning applies to this wider class. }.

\bigskip

Andreka and Nemeti tell us that the above mathemartical results (which were thought to be well established) are
{\it context dependent; they depend on the physical nature of our world.
They are an outcome of a Newtonian world, a world in which time is absolute.
By changing the underlying laws of physics, it is consistent with general relativity that one can
compute non-computable functions in the older \lq\lq Newtonian\,\rq\rq \, framework,
and one can prove $ZFC$ consistent}.

\subsection {Gravitational Radiation}

In electromagnetic theory, the acceleration of a charged particle produces electromagnetic radiation (that is, light, radio waves, $X$-rays, and so on). One
can think of electromagnetic radiation
as \lq\lq ripples\rq\rq \,in the electric and magnetic fields which propagate through spacetime at the speed of light.

\bigskip

In general relativity in a nearly flat region of spacetime one finds closely analogous behavour. Acceleration of a mass produces gravitational radiation
which can be thought of as ripples in the spacetime curvature, again propagating at the speed of light. In a {\bf strongly} curved region of spacetime the
distinction between these ripples and the background curvature of spacetime is unclear and one cannot give a precise definition of
gravitational radiation \cite{a}.\footnote{This is one of the difficult technical problem in general relativity \cite{a}.}\,
However, outside a strongly curved region (for example, far away from a black hole) the notion of gravitational radiation is unambiguous.
Astrophysical events which are likely to produce large amounts of gravitational radiation are

\begin{description}
\item [(a)] Supernovas or other gravitational collapse phenomena.\footnote{A supernova should produce a large burst of neutrinos.}

\item [(b)] Accretion of a star into a black hole at the center of a galaxy or star cluster.

\item [(c)] Coalescence of neutron stars and/or black holes in a close binary orbit. Hence, the detection of a burst of gravitational
radiation may give us information about phenomena probably involving a black hole.

\end{description}

%\newpage

\subsection{The Twin Paradox}

According to the first postulate of the special theory of relativity, the velocity of light is the same for all observers.
From the constancy of the speed of light follows the relativity of time. It is not true that there is an \lq\lq objective time\rq\rq \, spanning
the whole universe.
{\it \lq\lq Time is not an absolute physical quantity flowing by itself with no relation to anything external, only due to its internal structure,\rq\rq}
as Newton believed, {\it but varies from one inertial frame to the other.} One and the same event, when observed from different frames, may appear to
occur during different time intervals. This is exactly what is meant by relativity of time.
In particular, the time of an event measured by an observer
$S$ relative
to which the event is stationary, called {\bf proper} time,
would be {\it less} than that measured by another observer $S'$ moving relative to the same event. Every particle
has its own {proper} time as measured by an observer moving with it. Moreover, the time
measured by the observer
moving with the particle (relative to which the particle is at rest)
would be {\it less} than that measured by other observers (relative to which the particle is moving).\footnote{More
specifically, let $k$ and $m$ be two inertial observers with relative constant velocity $v$.
Suppose that $e_1$ and $e_2$ are two events occuring at the {\it same} location with respect to $k$.
Then, owing to the motion of $m$ relative to $k$, the two events necessarily occur at {\it different} locations with respect
to $m$. Let $t_k = time_k(e_1, e_2)$ and $t_m = time_m(e_1, e_2)$ be the time measured between the two events by
both $k$ and $m$, respectively. Then $t_k$ is the {\it proper} time (the two events occured at the {\it same} location
relative to $k$), $t_m$ is the {\it coordinate} time (the two events occured at {\it distinct} locations relative to $m$) and
$t_k$ is {\it strictly} less than $t_m$; such a difference increasing with the increase of the magnitude of the
relative velocity $v$. In fact, using the constancy of the speed of light, Einstein presented the following ingenous
thought experiment to show that time is not absolute
but varries from one frame of reference to the other. He argued as follows: We take $k$'s frame of reference
to be a train moving with uniform velocity $v$ relative to $m$ who is standing
on the platform. Observer $k$ holds a torch in his hand. The ceiling of his carriage has a mirror attached to it. He
turns on his torch sending out a light signal towards the mirror. Let $e_1$, $e_2$ be the sending and the reception of the light signal, respectively.
Relative to $k$, the light ray is reflected
back from the mirror reaching him at the same location from where he has sent it. Assuming
that the mirror is at a perpendicular distance $d$ from where $k$ is sitting, the light ray, according to $k$,
has travelled the distance $d_k = 2d$ going vertically up and down. The two events occured at the same location with respect to $k$. On the other hand, because the
train is moving relative to $m$, the sending and the reception of the light ray take place at
{\it distinct} locations according to him. This implies that the path traversed by the light ray is
{\it larger} than $2d$. In fact $d_m = 2(d^{2} + \frac{1}{2} (vt_m)^{2})^{1/2}$. Indeed, the light ray
traverses a \lq\lq traingular" path with base $z = vt_m$ and height $d$. Since the speed of
light is the same for both observers, it follows that $t_k = d_k/c < t_m = d_m/c$. In fact, a very
elementary calculation based on Pythagoras theorem shows that $t_k = (1 - (v/c)^{2})^{1/2}t_m$. {\it One of
the most important and revolutionary ideas in the history of science, namely, the
relativity of time, is deduced by Einstein from a very simple thought experiment that can be
easily understood by a teenager!}}
In brief, proper time is the time registered by a particle using its own clock. All observers agree on the proper time. Proper time, unlike
\lq\lq coordinate time\rq\rq, is an absolute physical quantity; {\it it
measures the rate of aging of the particle in motion: the faster the particle moves, the slower it ages! Accordingly,
clocks associated with moving
particles are slowed down and so a photon of light does not age!}\footnote{The length of an object also looses its absolute nature.
{\it Moving objects tend to contract in their direction of their motion}. Any object, for example, a rod of {proper length} $l_o$ (its length in its rest state.)
would appear to have length $l$ less than $l_o$ when viewed by
another observer relative to which it is moving; such a contraction depending on the (square of the) velocity. In fact, $l = (1 - (v/c)^{2})^{1/2}l_o$. Consequently, the distance between
points in space
and the interval between points in time are not, as Newton claimed, well defined unambiguise concepts, but depend on the position of the observer. As
Einstein says \lq\lq {\it It is neither the point in space nor the instant in time, at which something happens, that has physical reality, but only the event
itself. There is no absolute relation in space, no absolute relation in time, but only absolute relation in space and time.}\rq\rq}

\bigskip

All kinds of objections were raised against the special theory of relativity due to its anti-intuitive
character and revolutionary ideas.
One of the earliest and most persistent objections centered around what was referred to as
the {\bf twin paradox.} This paradox was in fact introduced by Einstein himself in 1905, in his paper on special relativity.
The twin paradox has caused the most controversy -
a controversy which raged on and off for over $50$ years. The paradox is usually described as a
thought experiment involving twins. Both twins
synchronize their watches. One of them then gets into a spaceship and makes a long trip through space.
Assuming that he has travelled with
a speed comparable to the speed of light time dilation would be large. Accordingly, he finds himself
{\it much younger} than his twin brother who has {\it stayed
behind on earth} as he has aged much less than his twin brother.
In fact, if the spaceship travels
just under the limiting speed of light, time within the spaceship will proceed
at a much lower rate; judged by the earth time (the stay-at-home twin's clock),
the trip may take more than one thousand years, whereas judged by the
travelling twin's clock, the trip may take only a few decades!
(the faster the spaceship, the less the travelling twin ages!) This involves
no paradox; being
actually a direct consequence of the predictions of the special theory of relativity.
The paradox, however, becomes apparent when we
describe this thought experiment
from the \lq\lq percpective\rq\rq \, of the general theory of relativity. Indeed,
according to the general relativity, there is no absolute motion of any sort; no preferred
frame of reference. It is always possible to
take any moving object as a \lq\lq fixed\rq\rq \, frame of reference.
This seems to suggest
that the situation is totally {\it symmetric}.
By taking the travelling twin as the fixed frame (now the earth makes a long journey away
from the ship and back again),
we are led to conclude that the {\it stay-at-home twin is necessarily younger than the travelling twin!}
{\it However, there is an important difference between the status of the two twin brothers; the stay-at-home twin
moves on a geodesic
of spacetime while the travelling twin does not}. Thus the situation is {\it not} symmetric after all.
One concludes from
the above arguments that the twin paradox, though anti-intuitive, is only an apparent and not a genuine
paradox. Morever, it is perfectly compatible with the predictions of both the special and the general
theories of relativity.

\subsection {A Glimpse of Einstein's Field Equations}

In a nutshell, Einstein's field equations say that
{\it matter curves spacetime and curved spacetime tells matter how to move.}\footnote{Accordingly, Einstein
has specified the mechanism by which gravity is transmitted: the wrapping of spacetime. Einstein tells us that the gravitational \lq\lq pull\rq\rq\,  holding
the earth in orbit is not, as Newton claimed, a mysterious instantaneous action of the sun; rather, it is
the wrapping of the spatial fabric induced
by the sun.} In more detail, Einstein's equation can be read as follows: matter -
represented by the energy momentum tensor - curves
spacetime (such curvature given by the Einstein tensor) and curved spacetime tells matter how to move,
namely, on timelike goedesics of the resulting geometric structure of spacetime.\footnote{In curved spacetime,
bodies having non-zero rest mass move on paths that {\it maximize} proper time choosing
the \lq\lq straightest" or \lq\lq longest" path compared to
all nearby paths. Such a property was described by the English philosopher and mathematician
Bertand Russel as cosmic \lq\lq laziness". In fact, geodesics in curved four dimensional spacetime are the
analogue of straight lines in flat spacetime. Geodesics appear to us as curved paths (for example,
the trajectories of planets around the sun) because
we tend to seperate between space and time. {\it Had we been four dimensional creatures, we would
\lq\lq see" geodesics as the straightest paths.}} Einstein's equation
representing such an interaction is prominantly {\it non-linear}:
matter distorts spacetime geometry, the resulting (distorted) geometry
constrains the motion of matter,  which, in turn, affects the geometry and so on and so forth.
The left hand-side of Einstein's equation is a purely {\it geometric} entity, whereas the right
hand-side is more of an emperical or {\it physical} nature. The equation is {not} to be regarded as
an {identity}; the Einstein's tensor does {\it not} define the energy momentum tensor. Rather, the equation
represents a {dynamic feedback loop} between two {distinct} yet {compatible} entities. This {dynamic}
relation between matter and geometry; this {feedback loop}, is {coded} in the following very elegant
(and simple!) equation
$$G_{ab} = 8\pi T_{ab}.$$
On the right, stands the {\bf source} of curvature, namely, the energy-momentum tensor.
On the left, stands the {\bf receptacle} of curvature
in the form of what one wants to know, the metric coefficients twice differentiated (the Einstein tensor).
The equation is in line with Mach's principle as expressed in $M_1$; the matter distribution
$T_{ab}$ determines the geometry $G_{ab}$, and hence is a source of inertial effects. Here we are
reading the field equations from right to left. We want to determine the metric coefficients from a given energy
momentum tensor, that is, the spacetime geometry corresponding to a given distribution of matter. Conversely,
we may regard the field equations as defining an energy momentum tensor corresponding to
a given spacetime geometry. In this case, we are reading the field equations from left to right.
It was originally thought that this is a productive way of determining energy-momentum tensors. However,
this rarely turned out to be very effective as the resulting energy-momentum tensors violated
some essential physical constraints.

\bigskip

When Einstein had created his general theory of relativity, he is supposed to have said that while the left hand side had been curved in marble, the
right hand side was built out of straw\cite{FDP}. The left hand side of Einstein's equations
refering to the actual geometry of spacetime is surely one
of the great insights of science. The right hand side describing how the mass and energy produces this curvature
did not follow with such elegance as the geometric part of the field equations.\footnote{Einstein always
hoped to extract the matter content of spacetime from its geometrical properties.
He always regarded the fact that the right-hand side of his
famous field equations containing the phenomenological tensor $T_{ab}$, an essentially non-geometric entity,
a blemish on his theory. In fact, Einstein was rather sceptical about the full field equations and
regarded the vacuum field equations, namely $G_{ab} = 0$, as more fundamental.} Most physical theories
nowaday, namely, string theory,\footnote{One of the
most bizzare and dazzling
predictions of string theory is that black holes which originally evolved in the context of general relativity
and elementary particles which
are quantum mechanical in nature {\it are one and the same thing}! In fact, string theory explicitely
establishes a direct, concrete and quantative
connection between black holes and elementary particles. Werner Israel, Richard Price, Stephen Hawking,
Roger Penrose and others have shown that
\lq\lq black holes have no hair\rq\rq;\, an expression invented by John Wheeler.  By this,
Wheeler meant the following: Any two black holes
with the same mass, force charge, and spin are completely {\bf identical}! This means that these three parameters
{\bf uniquely} determine a black hole and nothing else.
It is exactly such properties that characterizes an elementary particle.
For a deep exposition of the monumental success of string theory, we refer to (\cite{Brian}, \cite{FDP}).}  twister
theory\footnote{Twistors,
in contrast to string theory, are essentially the work of one man, Roger Penrose. The origins of twister theory lies,
not in the main stream of particle
physics, but rather in the field of relativity theory and the mathematics of complex spaces.
Yet, like superstring theory, twisters also deal with gravity, quantum
mechanics, and the nature of elementary particles. More precisely, the twister approach
suggests ways of relating gravitational curvature to quantum
mechanical transformations: A \lq\lq wave of curvature\rq\rq\,\, in spacetime looks like a \lq\lq quantum process\rq\rq\,
in twister space. Conversely, a quantum transformation in
twister space looks like a wave of curvature in spacetime. In other words, in the rich arena of twister geometry,
quantum and relativity theories
may be viewed, in some sense, as being {\bf dual} to each other! This suggests that there
may be a deep connection between general relativity, gravity
and spacetime on the one hand, and quantum processes on the other. It even suggests that one
theory may be affected by the other. In particular,
if one takes the nonlinear aspect of gravity into consideration,  then the logical core of the quantum theory, a linear theory,
may be altered! Most physicists believe that general relativity has
no influence on the logical structure of quantum mechanics. Penrose, however, holds the
unconventional view that the logical structure of the quantum theory
needs to be modified and this could only happen if general relativity is taken into account.
A detailed non-technical discussion of twister space may be found in \cite{FDP}.}
... are attempts {\it to fully understand the {right} hand side of Einstein's field equations trying to
establish a discrete description of (the geomerty of) {space and time} that might lead to a
smooth merging of both quantum and relativity theories}.\footnote{If we keep the focus on the
attempt to unify relativity and quantum theory, then we are continually impressed by the fact that
each of these are transitional theories. Each radically challenges the Newtonian conception of the Universe,
but only in part. Each holds unchanged a
certain, but different, part of the classical picture.
Relativity denies the absolute nature of the Newtonian world, retaining its deterministic
aspect, while quantum theory abolishes the
deterministic aspect of the classical world, retaining the absoluteness of the Newtonian reality.
So the situation is genuinely confusing. Despite more than half a century of hard work by
some of the world's leading physicists, these two theories have
stubbornly refused to be reconciled, but continue to co-exist in paradoxical and incompatible ways.
The notion of a smooth spatial geometry, the central
principle of general relativity, is destroyed by the violent
fluctuations of the quantum world on short distance scales. On
ultramicroscopic scales, the central feature of quantum mechanics -
the uncertainty principle - is in direct conflict with the central
feature of general relativity - the smooth geometric model of
spacetime.}

\bigskip

We note that the field equations show how the stress energy of matter generates
an average curvature in its neighbourhood. It governs
the external
spacetime curvature of both a static and dynamic source,
the generation of gravitational waves (ripples in the curvature of spacetime) by stress energy
in motion, the external (and internal) spacetime geometry of a (static and rotating) black hole and,
last but not least, the expansion and the contraction of the universe.

\bigskip

In conclusion, the field
equations not only described the dynamics of the universe on large scale, but also
predicted mind-boggling physical phenomena (black holes, event horizon, time wraps, gravitational waves,
\ldots etc) that opened up new horizons for research in the world of physics;
phenomena that are till this
very moment (almost a century after the discovery of general relativity) the subject of intensive
investigation and thorough scrutiny!

%\newpage

\end{mysect}

\begin{mysect}{An Axiomatic Approach to Relativity Theories}

Logical axiomatization of physical theories is far from being a new idea.
It goes back to such leading mathematicians and philosophers as Hilbert, Godel, Tarski, Reichenbach, Carnap, Suppes, \ldots and
many other eminent scientists. There are many examples showing
the benefits of such an axiomatic approach when applied to the foundation of mathematics. Accordingly,
it is natural and useful to apply such a method to physical theories; in particular to
spacetime theories like both the special (SR) and the general (GR) theories of relativity.

\bigskip

According to the English philosopher and mathematician Bertand Russel,
{\it mathematics does not tell you what is but what will be if,
a physical theory is supposed to tell you what is.}\footnote{This is only true in
the realm of {\it classical} physics. In the micro world,
matters are more complicated; what we observe is not nature itself, but rather nature as revealed to us through
our methods of questioning. The last part of this paper discussing Einstein's view on the
quantum theory will tackle such issues.} Accordingly, the
process of axiomatization in the realm of mathematics and physics
serves different purposes.\footnote{In mathematics, any theorem that we prove is correct
in an {\bf absolute} sense.
According to the Austrian philosopher Karl Popper (regarded as one of the
most important philosophers in the twentieth century), the nature of physical knowledge is different. {\it We can't
prove the \lq\lq correctness" of a physical theory}. In fact, any physical theory is necessarily {\it incomplete}.
It is valid only in a {\it certain restricted domain}. The {\it strength} of a
physical theory is measured by its {\it informative content}. The more
it tells us about the external world, the more it is powerful and the more it can be potentially \lq\lq proved wrong".
Accordingly, physical theories can only be {\bf falsified} and {\it not} verified. A single phenomenon that escapes
the explanatory power of some physical theory indicates that the theory is
necessarily a special case of a more general one that captures such a phenomenon and
enlarges the domain of applicability of the original theory. Progress in the
physical world is thus accomplished by the replacement of \lq\lq weaker" theories by \lq\lq stronger" ones, that is,
theories that have more \lq\lq informative" content. {\it To sum up, the notion of {\bf falsification} states that
our physical knowledge develops by the overthrow of hitherto well-established
theories to be replaced by \lq\lq better\rq\rq ones, that is theories that
have a larger and wider scope thus have more explanatory power. In the language of Darwin, physical theories compete and the ones
surviving are the fittest!}}
{\it In the world of mathematics,
we are free to choose any set of axioms;
statements that we assume without proof,
as long as they are non-contradictory}. We don't
care whether such axioms or their logical consequences
describe any properties of the external world. What really concerns us is the
compatibility of the chosen axioms and their
consistency (their independance is also important for aesthetic considerations).
{\it On the other hand, the status of axioms in the world of physics is
different. We are not totally {free} in our choice of the axioms
since our aim is to describe certain regularities
of nature. As such we are constrained by the laws imposed on us by the outside world. The
domain of application of our chosen axioms is somewhat limited and in some sense fixed.}

\bigskip

The role played by
the axioms in the formulation of a physical theory is thus more intricate than their role in the
formulation of a mathematical theory.  {\it The axioms
chosen for a physical theory should be simple, logically transparent, illuminating,
intuitively clear and easy-to-believe. All surprising, bizzare. exotic, unexpected or unusual predictions of the
theory are to be described, not by the axioms themselves, but rather by
theorems derived from the axioms (which we may refer to as "fancy" theorems).}\footnote{For
example, in the context of SR, the property that no inertial observer moves faster or
equal to the speed of light should be derived as a theorem and not be put as an axiom.}

\bigskip

The process of axiomatizing a physical theory
provides both precision and rigour for the theory.
In fact, the {\it utility} of the axiomative system chosen to describe a physical theory could
be measured by its ability to
describe {\it exotic} and strange phenomena from {\it clear} and
self-evident axioms. Another important advantage that could be gained from the axiomatization of a physical theory
is to be able to derive most, if not {\it all} the interesting and unpredictable phenomena of the
theory from the {\it smallest} number of {independent} and simple axioms.

\bigskip

Both theories of relativity have consequences that are anti-intuitive and defy common sense. Most of their
predictions are bizzare and far from being evident. For these reasons, it is interesting and
important to try to construct an axiomatic approach to these spacetime theories. This might make us
better comprehend the exotic phenomena predicted by both theories and throw light on the meaning of the apparently
paradoxical predictions occuring in them. {\it In fact, by {playing} with the axioms,
such an approach may actually lead to the prediction of physical phenomena that transcend the
domain of applicability of these spacetime theories. Indeed, one of the aims of axiomatizing a physical
theory is not only to reveal its physical content, but also, if possible, to enlarge its scope and its power of
predictability.} For example, by weakenning or adjusting the axioms describing GR, we
might be able to throw light on the logical structure of the more general five-dimensional
Kaluza-Klien theory unifying gravity and electromagnetism on a geometric basis.\footnote{Postulating
the existence
of an extra curled up space dimension, the Polish mathematician Theodor Kaluza argued
that gravity is carried by ripples in the familiar three space dimensions, while electromagnetism
is carried by ripples involving the new, curled up dimension. The reason for the unobservability
of the fifth dimension (its compactness) was suggested by the Swedish physicist Oskar Klien. Today
the theory is called the $5D$ Kaluza-Klien theory. (Kaluza sent his paper to Einstein in 1919. Einstein
regarded Kaluza's assumption of the existence of \lq\lq an extra curled up space dimension"
not very convincing. Kaluza did not publish his paper until two
years later when Einstein became more at ease with Kaluza's assumption giving him the
green light to publish it. In fact, Kaluza's \lq\lq wilde" idea was way far beyond his time; namely,
\lq\lq unifiying" apparently distinct physical phenomena by \lq\lq jumping into a new dimension". It
seemed \lq\lq crazy" even for Einstein himself. Though it
failed experimental tests, this idea of postulating the existence of extra curled
up space dimenstion was to be revived again in the context of (super)string theory. Modern versions of
the Kaluza-Klien theory go up to ten space dimension (six of which are curled) and
one time dimension. These \lq\lq extra degrees of freedom" make such theories
have the potential flexibility to merge all known forces of nature (including gravity!) into one harmonious
framework \cite{Brian}}. This may
even open the door for dealing with other physical theories having more
informative content and stronger explanatory power. {\it One possible way to accomplish this is that the questions adressed when investigating the logical structure of relativity
theories should not be only the \lq\lq how" questions but, more importantly, the \lq\lq why" questions: What is
believed and why? Which axioms are responsible for what predictions? What happens if we
weaken some of the axioms? Can we change (some of) the axioms and at what price?}

\bigskip

In their attempt to axiomatize GR, Andreka and Nemeti
first present  streamlined axiomatic system for SR. From this simple and naturel set of axioms all paradmatic
effects of SR are derived: moving clockes slow down, moving rods shrink and
moving pairs of clocks get out of synchronism\ldots etc, thereby capturing
all exotic predictions of SR. The transition from
SR, which is based on the notion of {\it inertial} obsevers, to GR is partially accomplished by
introducing the more general notion of {\it accelerated} observers. In the context of
accelerated observers, the twin paradox is
derived as a theorem being a logical consequence of the more flexible set of axioms
describing accelerated observers. {\it The elimination of the difference between inertial and accelerated observers
{\it on the level of axioms} leads to an axiomatization of the spacetime of GR.}
In fact, using the equivalence principle as a guide, the axioms chosen for
GR are in a (well defined) sense a {\it localization} of the axioms describing SR.
The models of GR are locally Lorentzian smooth manifolds in which
every local chart is equipped with a locally Minkowskian metric.
\footnote{Technically, a topological manifold $M$ of dimension $n$ is a topological
space that is Hausdorff (distinct points can be seperated by disjoint open sets)
and locally Euclidean: every point $p\in M$ has a neighbourhood
homeomorphic to some open subset of $R^{n}$; i.e., for each $p\in M$, there exists a pair
$(U, \phi)$ where $U$ is open in $M$ and $\phi:U\to \phi(U)$ is a homeomorphism (both $\phi$ and its inverse $\phi^{-1}$ are continuous). The
pair $(U, \phi)$ is called a {\bf local chart}. We say that $(U, \phi)$ and $(V, \psi)$ are
$C^{\infty}$ {\bf compatible} if whenever $U\cap V \neq \phi$, then the transition function
$\psi\circ \phi^{-1}:\phi(U\cap V)\to \psi(U\cap V)$ and its inverse are $C^{\infty}$ as mappings of open subsets of $R^{n}$.
A $C^{\infty}$ differentiable (smooth) structure on $M$ is a family ${\cal U} = (U_i, \ \phi_i)_{i\in I}$, with
$U_i$ open, such that (a) the $U_i's$ cover $M$. \ (b) the $(U_i, \ \phi_i)_{i\in I}$ are $C^{\infty}$ compatible. \
(c) ${\cal U}$ satisfies the
following {\bf maximality} property: every $(U, \phi)$ which is $C^{\infty}$ compatible with each and every element
of ${\cal U}$ belongs to ${\cal U}$. A smooth manifold is a topological
manifold equipped with a smooth structure.} Roughly, these locally Lorentzian manifolds $M_L$\footnote{For physical considerations, $M_L$ is not necessarily Hausdorff.} are
constructed as follows:  The points of $M_L$ stand for the events; an event being nothing but a set
of bodies, $I$ is an indexing set
numbering the observers, $U_i$ stands for the frame of reference of the observer $i$ and,
finally, the transition functions $\phi_{ij}:U_j\to U_i$ are the world view transformation between
the $i$-th and $j$-th observers. Each local chart $U_i$ is equipped with a (local) Minkowski metric $g_i$.
These local metrics $g_i's$ are then \lq\lq lifted" to the whole manifold by constructing the
tangent space $T_e(M)$ at each event $e$.\footnote{Having the notion of a tangent space,
the concept of a tensor can be introduced in the axiomatic language of GR; a multilinear map
of the cross product of the
tangent space with itself and its dual to the field $Q$. The field $Q$
plays the role of the real numbers $R$, namely representing the "quantities". The axioms satisfied by
$Q$ are {\it weaker} than those satisfied by the real field $R$.
In fact, the "quantity" part $(Q, +, ., <)$ is a Euclidean ordered field; an ordered field in the
sense of abstract algebra in which every {positive} number has a root.} The process of
\lq\lq lifting" these local metrics produces a {\bf global}
metric defined on the entire Lorenzian manifold.

\bigskip

The three theories - special relativity,
accelerated observers and general relativity - are
formulated in first order logic (FOL). There are at least two reasons for this. First, to avoid any
tacit assumptions so that all concepts dealt with are clear and well defined.
Secondly, by Godel's incompleteness, FOL has a complete inference system which
higher-order logics do not  have.

\end{mysect}

\subsection {Breaking the Turing Barrier}

We here give a very concise summary of the construction of relativistic computers breaking the Turing barrier.
Using Malament-Hogarth spacetimes and other general relativistic phenomena, Andreka and Nemeti
have succeded to construct (what they refer to as) {\bf relativistic} computers.
Relativistic computers are based on the property of time dilation. As previousely mentioned, among the interesting and
bizzare predictions of the general theory of relativity is that
time slows down in strong gravitational fields.\footnote{This can be actually deduced
from the postulates of the special theory of relativity together
with the equivalence principle: The special theory of relativity tells us that a clock associated with
an accelerated observer slows down. Since gravity and acceleration are indistinguishable
(the equivalence principle), it follows that
clocks tick slower in a gravitational field.}

\bigskip

A \lq\lq stationary\rq\rq\, (Schwarzchild) black hole has \lq\lq one event horizon\rq\rq. The event horizon acts as a
{\it one-way} membrane surface; one can pass through the horizon, but once inside one can never leave.
The event horizon represents the {\it boundary} of all events which can be observerd by an
{external} observer; if one crosses this boundary, one becomes trapped behind it. The
Schwarzchild event horizon is {\it absolute} because it seals off all internal events from every external
observer. A {\lq\lq rotating\rq\rq\,} black hole, unlike a stationary one, has \lq\lq two event horizons\rq\rq.
The gravitational pull of stationary (rotating) black holes grows without limit
as one approaches the (outside) event horizon.

\bigskip

Assume now that two observers, which we denote by $H$ and $L$ respectively, are hovering near
the outside event horizon, with $H$ being
higher up. Then $K$'s clock runs slower than $H$'s clock since he is experiencing a stronger gravitational field.
Moreover, as $L$ moves towards the horizon, this discrepancy between the ticking of both clocks gets
larger and larger. In fact, by lowering $L$
appropriately, we can actually control this \lq\lq time lag\rq\rq. Now, if a programmer $P$ gets
very close to the outside event horizon while leaving his
computer $C$ \lq\lq higher up\rq\rq, then in a few days time relative
to the programmer, the computer does a few million's year's job! Accordingly,
one can reach an \lq\lq infinite speed up\rq\rq\,   by lowering
$P$ to the right position; hence breaking the \lq\lq Turing barrier\rq\rq. The above mentioned
thought experiment, however, cannot be carried out in a
Schwarzchild, that is, a non-rotating black hole.
This is because either $L$ will be destroyed from the gravitational might or
some photon sent by $H$ would not reach him. A possible way out is to choose a
slowly {\it rotating} black hole. The rotation of the black hole induces a
repelling effect - a centrifugal force in the langauge of Newtonian mechanics,
that counter-balances the strong gravitational pull of the black hole.
In this way, $L$ can slow down as desired without being crushed.

\bigskip

As previously stated, a slowly rotating black hole has two event horizons. The outer one is similar to
that of a Schwarzchild black hole. The
{\it inner} event horizon  is the one that {\it overcomes} the gravitational force. Accordingly,
it is possible for an observer $L$ to stay at a fixed distance from the
center of the rotating black hole. In this way, an infinite time dilation (speed-up)
of the computer $C$ which lies safely away from the (outer) event horizon
with respect to the programmer $P$ is accomplished. The creation of a computer that can
compute tasks beyond the Turing limit can be
achieved as follows.\,The programmer $P$ leaves earth in a
spaceship towards a huge slowly rotating black hole. As $P$ is heading towards his target,
$C$ checks one by one the theorems of set theory. If $C$ finds a contradiction,
he sends a signal to $P$. Otherwise, he does nothing. Now, what happens to the
programmer $P$ from $C$'s point of view. As the programmer $P$
approaches the event horizon, his clock will be ticking slower and slower relative to $C$'s clock. At the {\it limit},
that is, when $P$ reaches the inner horizon, his clock {\it freezes}, coming to a halt, so to speak,
relative to $C$. From the point of view of $P$, however, the  $C$'s clock
appears to be running faster and faster. Moreover, assuming that the black hole is huge
so that the tidal forces on the event horizon of the black
hole are negligable, $P$ will safely cross the inner event horizon.
Two things can occur: either $P$ recieves a light signal from $C$ or not. In the latter
case, $P$ will know that $C$ has found an inconsistency in ZFC set theory.
Otherwise, $P$ concludes that $ZFC$ is consistent. Finally, why the choice of a huge
rotating black hole? There are actually two reasons for this: first, because the black hole is huge,
the center of the black hole is relatively
far from the event horizon. Secondly, and more importantly, the matter content, that is,
the singularity is not a point as in the case of a static black hole, but is
actually a {\it ring}. This is one of the fascinating properties of rotating black holes. These two features
make $P$ comfortably pass through the
middle of the ring without being crushed or torn apart!

\bigskip

We end this part of the paper by posing the following question: Andreka's and Nemeti's exotic results (computing non-computable functions,
proving that $ZFC$ is
consistent, $\ldots$ etc) are actually obtained in the context of the {\bf classical} general theory of relativity.
These results are a consequence of the
property of infinite time dilation which occurs in the strong graviational fields in the vicinity of rotating black holes.
A natural question arises: Will these
results still hold if we take the {\bf quantum} theory into consideration. In this case, we will be
dealing with (what we may refer to as)
quantum relativistic black holes (QRBH). These behave
in a manner categorically different than classical (rotating) black holes.
For example, they emit radiation\footnote{According to Hawking, this occurs even if
the black hole is not rotating!} so they can actually glow. Will infinite time dilation
still hold in the framework of QRBH? And if not, will Godel be right after all?
If this is so, then there might be a way to relate the uncertainty principle (the corner stone of quantum theory)
with the Godel's undecidability (the corner stone
of mathematical logic) near a quantum black hole. Essentially, both principles are derived
from the notion of \lq\lq self reference\rq\rq: Heisenberg's uncertainty relation is a
consequence of the merging of subject (observer) and object (observed);
whereas Godel's undecidability results from a merging of subject (mathematics) and object (meta-mathematics).
In the classical context of black holes, in which infinite time dilation occurs, Andreka and Nemati have
shown that ZFC is consistent. In the quantum realm, that is, taking the uncertainty principle into considerations,
it is not quite clear that the property of time dilation still holds so that ZFC may again
be undecidable.  {\it Accordingly, the two most important {limitative} results discovered in the 20-th century may
be an outcome of a more fundamental law describing the true physical nature of quantum black holes!}

%\newpage

\begin{mysect}{On the Meaning of the Quantum Theory}

{\it One does not get an answer to the question, What is the state after an atomic collision? but only to the question, How probable
is a given effect of the collision? From the standpoint of our quantum mechanics, there is no quantity which causally fixes the effect of
a collision in an individual attempt. Should we hope to discover such properties later \ldots and
determine them in individual events?
\ldots I myself am inclined to renounce determinism in the atomic world, but this is a
philosophical question for which physical
arguments alone do not set standards.} Max Born as quoted in \cite{a}.%Lee

\bigskip

{\it In a way, quantum mechanics reminds us of the old wisdom that when searching for harmony in life, one must never forget that in the drama
of existence we are ourselves players and spectators.} Neils Bohr.

\bigskip

{\it The spacetime continuim may be considered as contrary to nature on view of the molecular structure of
everything which happens on a small
scale..... Perhaps the success of the Heisenberg method points to a purely algebraic method of
description of nature, that is the elimination
of continuous functions from physics\ldots  At the present time, however, such a program looks
like an attempt to breath in empty space.} Albert Einstein, Out of My later Years.

\bigskip

{\it My own view is that ultimately physical laws should find their most natural expression in terms of essentially
combinatorial principles, that is to say in terms of finite processes such as counting.\ldots Thus, in accordance
with such a view, should emerge some form of discrete or combinatorial spacetime.} Roger Penrose,
Magic without Magic

\bigskip

{\it There is nothing wrong with the uncertanty principle. Einstein was simply confused}. Stephen Hawking

\bigskip

{\it I believe that no one truly understands quantum theory.} Richard Feynman.

\bigskip

Chance and necessity, cause and effect, choas and randomness are words commonly used in today's world of physics reflecting
the abstruse nature of the micro-world. The seemingly paradoxical
laws of quantum mechanics, strangely enough, have given us a more exact model of our universe, in the sense that they have not only overcome
contradictions which remained unresolved in the realm of classical physics, but also predicted unknown and bizzare
phenomena giving us a radically new and deeper cognition of the world around us.

\bigskip

The statistical nature of the new-born theory at the turn of the century made most physicists sceptical towards it, believing that the theory was
incomplete, and hence part of a more general theory based on non-probabilistic laws. Einstein was never at ease with the uncertainty
principle, the cornerstone of quantum theory, expressing his doubts in his famous saying \lq\lq God does not play dice with nature.\rq\rq
\footnote{According to Stephen Hawking, God not only plays dice with nature, but throws them in places where you cannot find them!} Einstein was actually one of the great critics of quantum mechanics, believing in a deterministic reality based on exact objective laws.
Neils Bohr, the great Danish physicist, not only showed no critical attitude towards the new theory, but was also interested in its
philosophical implications, as proved by his \lq\lq principle of complementarity\rq\rq. The younger
generation, not yet prisoners of the well established theories and concepts of classical physics, were in a
somewhat uncomfortable
position. Baffled by the new discoveries on the one hand, and on the other, being more flexible than Einstein,
they were able to cope
brilliantly. \lq\lq Heisenberg's matrix-mechanics\rq\rq, \,\lq\lq Schrodinger's wave equation,\rq\rq \, and, finally,
Dirac's \lq\lq bra-ket
formulation\rq\rq of quantum mechanics\footnote{Technically, the mathematical framework of the
quantum theory is an infinite dimensional Hilbert space equipped with the delta-Dirac function.}, which
generalized and proved the mathematical equivalence of both approaches,
marked the end of the first period (old quantum theory, which started with the
discovery of the discrete nature of radiation by Max Plank) and was the beginning of a new era in quantum physics based on solid
mathematical foundations.\footnote{In the realm of classical physics, that is, in Newton's
and Maxwell's theories, the property of a \lq\lq wave\rq\rq\, and a \lq\lq\ particle\rq\rq\, are simply
mutually exclusive, representing \lq\lq incomensurable\rq\rq\, notions.
A wave (resp. particle) cannot exhibit corpuscular (resp. wave)
properties. Particles are
discrete, being localized in space, whereas a wave is continuous and extended in space. Light, according
to Maxwell's theory is unquestionably a (special type of an electromagnetic) wave. It is subject to
both interference and diffraction phenomena, properties that are undeniably allian to
the behavour of particles. On the other hand, particles as described in the Newtonian framework do not reveal any
wave-like nature. It was only at the turn of the 20-th century that Einstein assumed a strange duality in
the behavour of light in order to understand the phenomena of the photo-electric effect.
To explain the emission of electrons from
the surface of (some kinds of) metals when being bombarded by light, Einstein postulated that
light has a \lq\lq grainy nature\rq\rq\, and
is composed of discrete particles which he called \lq\lq quanta\rq\rq.\, In doing so, Einstein was
actually generalizing Plancke's ad-hoc assumption proposed five years earlier on the nature of (electromagnetic)
radiation. Contrary to the classical Maxwell's theory,
Planck had to assume that electromagnetic radiation is composed of discrete
units to explain the phenomenon of the
black body radiation in an attempt to resolve (what was known as) the ultra-violet catastrophy.
These ideas were further developed by the French physicist de-Broglie and set in a much wider framework.
An intuitive belief in
the \lq\lq symmetry\rq\rq\, of nature made
de-Broglie propose the following daring
assumption: not only light, under certain circumstances, may reveal corposcular nature (as
Einstein has rightly discovered), but all matter objects are associated by (what he called) matter waves. Thus
de-Broglie was actually extending the dualistic nature of light to all forms of matter! It was
a few years later, through the work of Bohr, Born, Schrodinger, Heisenberg, Dirac and others that such a
revolutionary conception was further elaborated in the probabilistic language of the quantum theory
resulting in a subtle form of {\bf duality} between
matter and radiation.}

\bigskip

{\it The essence of the problem is that quantum mechanics is essentially a more complete theory than classical
physics and yet
describes the universe in a statistical manner.} The idea of a statistical approach in describing natural phenomena
was, of course, already introduced
in the realm of classical physics, namely in the theory of thermodynamics, dealing with concepts like entropy
and free energy. However, the
meaning attributed to such a statistical analysis was of a categorically different nature than that of the
new-born quantum theory. It was
always assumed that, whenever probabilistic laws emerged in relation to any physical phenomenon in the
classical domain, the element
of indeterminism was purely due to a lack of knowledge; an incompleteness of the data necessary to
describe the phenomenon under study. Physicists
believed, at least in principle, that it was always possible to overcome or to eliminate the statistical aspect once one was able to gather all the
necessary information about the phenomenon in question. The lack of information, that is, the ignorance in relation to the system studied, is therefore
due to a defficiency in the observer (subject) and not in the observed (object). {\it In quantum mechanics,
matters are more subtle and intricate. Statistical
laws emerge from within, being an innate fundamental aspect of the theory. Having a probabilistic nature, quantum mechanics seems to give an
incomplete picture of the world; a fuzzy description of reality. By definition, probabilistic laws can only predict what might be and not what is.
When we talk quantum, the focus is on what we do and what we observe, rather than what is.}

In the micro world, \lq\lq subject and object\rq\rq, \lq\lq observed and observer\rq\rq \, are not totally separated. As a result
of this interaction, a kind of fuzziness evolves, an indeterministic factor that seems to dominate all micro phenomena. As
one moves from the small to the large, as one enters the world of the macro, where the subject no longer affects the object, this fuzziness
or blurdness gradually disappears. The uncertainty relation, according to the above argument, could be realized as the law that
measures the degree of such merging; equivalently as a kind of limit beyond which the line of demarcation between subject and object becomes
unclear and not well defined. However, this is not the whole story. Many questions remain unanswered,
the most important of which, we believe, is the following:
Is nature itself indeterministic, in the sense that the \lq\lq grey zone\rq\rq\,, an essential feature of the
quantum world, is intrinsic in the
universe, or is it due merely to our inability to know nature without meddling with it? In other words, does nature have a probabilistic
essence, or does it only reveal itself to us through non-exact laws? Though some may argue that
the above two questions are categorically different, we claim that the difference is of no relevance. Talking
about the nature of the universe as a thing in itself is simply meaningless. What counts is what we (the subject) can know about nature
(the object). We discover nature through ourselves. The laws of physics are defined only in relation to us, acquiring their meaning due
to our existence. They are not floating in the cosmos, so to speak, but are out there to be extracted by us. The process of this
extraction  makes the universe, dead as it may seem, hit back revealing itself in a somewhat blurred form. As Heisenberg says,
in his book, Philosophy and Physics: \lq\lq What we observe is not nature itself, but rather nature exposed to our methods
of questioning.\rq\rq

\bigskip

Quantum mechanics made a revolutionary impact on our views towards the world and even
towards ourselves. Scientists were forced
to alter their beliefs and dogmas that were inherited from past ages.
Words like \lq\lq knowledge,\rq\rq\, \lq\lq understanding\rq\rq\,
and other fundamental concepts acquired a completely new meaning.
Theories that were well established proved inadequate and incomplete,
being only an approximation of reality. Scientists learnt not to take anything for granted, becoming more skeptical in dealing with new concepts
and theories.
The fascinating, if not unbelievable, consequences of the new theory made physicists more daring in their imagination.
Any assumption, no matter how absurd, was admitted
as long as it was compatible and consistent with the axiomatic scheme of quantum laws. In less than five years, after the mathematical foundations of the
new theory were laid down, the progress achieved in the newly born quantum theory was greater than that achieved in classical physics
throughout the whole of the nineteenth century. It was a decisive period in the development of science; a
time in which second rate physicits could come up with first rate ideas. Wild and crazy interpretations accompanied the new science, like the \lq\lq many world
interpretation\rq\rq\, of quantum mechanics,\footnote{The world we see around us must be only a tiny part of reality. Instead,
the universe is multiplied into an infinite number of copies or \lq\lq parallel branches\rq\rq\, so that
there is a branch for each of the
possible outcomes of every experiment and observation. This astounding suggestion was made in the 1950's by
a graduate student named Hugh Everet. It was taken up and champoined by two of the great pioneers of quantum
cosmology, Bryce DeWitt and John Wheeler. The idea behind it was to eliminate \lq\lq the dicotomy between the
observer and the observed\rq\rq\, upon which is based the Copenhagen interpretation; making the observer part of the
universe.} the \lq\lq absence\rq\rq\,  of  cause and effect,
and hence the possibility of time flowing backwards.\footnote{Under this hypothesis, Richard Feynmann,
the great American physicist and one of the pioneers
of theoretical physics during the second half of the twentieth century, showed that a positron is
mathematically equivalent to an electron moving
backwards in time! In fact, Feynmann has even succeeded to invent a new
formulation of the quantum theory. In Feynmann's formulation of quantum mechanics, an object travells
from one location to another by
\lq\lq sniffing out all possible trajectories\rq\rq. The resulting motion observed is thus a combination
of all possibilities,
with the relative contribution of each possible trajectory
precisely  determined by the mathematics of quantum mechanics. Feynmann showed that he could then assign a
(complex) number to each of these paths in
such a way that their combined average yields exactly the same result for the probability calculated using the
ordinary wave function approach.
In other words, the probability that an object moves from one point to the other is given
by the combined effect of every possible way of getting there; instead of the particle {\it not having a well-defined
path} when moving from one location to the other (as in the \lq\lq conventional\rq\rq\, quantum theory),
in Feynmann's
approach, the particle {\it traverses all and every possible path between the two locations}.
This is known as Feynmann's
\lq\lq sum over paths (histories)\rq\rq approach to quantum mechanics.}
Words like \lq\lq ghost particles\rq\rq \, and \lq\lq tunnelling effects\rq\rq\, \footnote{For microscopic particles facing an energy
barrier, they can borrow enough energy to momentarily penetrate and tunnel through this region.
This is
a consequence of the uncertainty relation; one reading of which is that
{\it a particle cannot have precise amount of energy at a
precise moment of time}. This means, roughly speaking, that the energy of a particle can \lq\lq wildly\rq\rq\,
{\it fluctuate} as long
as this fluctuation happens in a short period of time; the energy being {undetermined} during this short
interval. The particle may then {\it momentarily borrow excess energy}, provided it pays it back
quickly within a time frame allowed by the uncertainty
principle. For microscopic particles facing a barrier, they can (and sometimes do)
use this \lq\lq borrowed energy\rq\rq\, to {\bf tunnel}
through a region which they did not initially have enough energy to enter; thus acting in a way
which has no analogue in the context of classical physics.} were introduced,
reflecting the abstruse and the (seemingly) paradoxical nature of the quantum world. The concept of a \lq\lq well-defined path\rq\rq \, was simply meaningless within
the framework of the quantum world. The \lq\lq wave particle\rq\rq \, duality that dominated all
atomic phenomena was given a
philosophical interpretation in Neils Bohr's \lq\lq principle of complementarity\rq\rq: \,
although the wave-particle properties
of atomic particles are
apparently \lq\lq mutually exclusive\rq\rq, \, they are, nevertheless, \lq\lq complementary\rq\rq \,\footnote{According
to the famous American philosopher, Thomas Kuhn - an analyist of the history of science, progess
of science is accomplished through the merging of apparently incompatible concepts through the langauge of {\it metaphores}.
Kuhn  claims that through the constructive and creative language
of metaphores, a dynamic process of interaction and communication would be
a guiding rule for the emergence of novel and daring ideas. These new ideas come into being, not through the
breakdown or downfall of older concepts, but rather by creating different
and wider contexts in which older theories may reveal original and unpredictable features. Concepts that
appear mutually irrelevent in one context may be combined, related, or even identified reflecting
one and the same phenomenon when viewed from a wider perspective. Kuhn says
that such a reconcilation of seemingly unrelated or even apparently contradictory phenomena has been
the mechanism by which the boundaries of scientific knowledge expand. To be sure,
it is the \lq\lq essence" of scientific develoment. For example, in the world of
the small, using the language of metaphores, one can say that
\lq\lq a wave {\it is} a particle". Another clear and illuminating example in favour of Khun's analysis is revealed in our
(classical) understanding of electromagnetism. Throughout the eighteenth century, physists
dealt with electric and magnetic effects in nature as two completely seperate phenomena described
by mutually irrelevent physical theories. {\it A partial unification of these two theories  was
first accomplished by Maxwell by discovering that a varrying electric (magnetic) field gives rise to
a varrying (magnetic) field; both together giving rise to the notion of an electro-magnetic
field whose evolution in space and time is mathematically described by Maxwell's equations.}
If Maxwell had succeded to show that a varrying electric and magnetic field co-exist together to
form one single entity, then Einstein,
through his theory of special relativity (SR), went a step further. {\it He was able to show
that one and the same field may be either
magnetic, electric or a combination of both when viewed from different \lq\lq angles"; the nature of
the field revealed being dependent on the state of
motion of the (inertial) observer studying it.} Accordingly, SR, not only showed a complete symmetry
between electric and magnetic fields - a fact already recognized within the context of Maxwell's theory,
but also established a unification of two fragments into one whole through the
metaphore \lq\lq Electricity is magnetism and magnetism is electricity.}
as opposed to \lq\lq contradictory\rq\rq. \,
An understanding of the nature
of the electron could be only achieved by combining both aspects, the two together constituting the essence of the electron. In one context
(experimental framework), electrons act like particles, while in another they reveal wave-like properties. In general,
the \lq\lq wave-particle\rq\rq \, duality was a reflection, in the language of old physics, of a deeper conflict
between the goals of \lq\lq description\rq\rq\, and \lq\lq causality.\rq\rq\,
One can describe the world at any instant to any desired accuracy and produce a \lq\lq snapshot,\rq\rq\, so to speak,
showing where
everything is. The principle of complementarity states that such a \lq\lq snapshot \rq\rq\, could be only
taken at the expense of {\it forewearing} any
connection between its future \lq\lq snapshots\rq\rq.\, The {sharper} the snapshot, the {looser} its causal
lies with the future. {\it We must
choose some compromise between an orderly, causal world which we cannot even visualize and a
sharp picture which reflects only the
instant it was taken.}

\bigskip

\bigskip

The abstract formulation of the new theory made it difficult to comprehend, even among physicists themselves. Based completely on pure
mathematics, physicists realized the impossibility of \lq\lq visualizing\rq\rq \, the world of the small. Resorting to pictorial images
in trying to understand atomic phenomena proved impossible. The meaning of the word \lq\lq understanding\rq\rq\, had to be revised and
defined in more general terms. If \lq\lq picturial images\rq\rq\, were meaningless in dealing with the atom, could mathematical rigour with
its inner consistency be an alternative? And if so, does this mean that a \lq\lq physical theory\rq\rq \, is nothing more than its \lq\lq
mathematical model\rq\rq\,? Paul Dirac, wrote in this respect: \lq\lq The main object of physical sciences is not the provision of pictures,
but is the formulation of laws governing phenomena and the application of these laws in discovering new phenomena. In the case of atomic
phenomena, no picture can be expected to exist in the usual sense of the word picture, by which is meant a model functioning essentially
on classical lines. One may, however, extend the meaning of the word picture to include {\it any way of looking at the fundamental laws
which make their consistency obvious}. With this extension, one may acquire a picture of atomic phenomena by becoming familiar with
the laws of quantum theory.\rq\rq

\bigskip

One of the interesting philosophical implications of quantum physics is the concept of \lq\lq free will\rq\rq.\, In the seventeenth century,
it was believed that the universe is totally deterministic, a kind of gigantic clock, subject to Newtonian mechanics. Pierre Laplace, a French
mathematician and physicist, reflected this belief by stating that if it were possible to express the equation of motion of every particle in the universe
at some time $T$, then the state of the universe at any other time $T$, by virtue of Newton's laws of motion, would be completely determined.
The present is, therefore, a mirror reflecting both the past and the future. This bizzare assumption was certainly a blow to the concept of
free will. We, as a part of this gigantic clock, operated according to mechanical deterministic laws, and are hence totally \lq\lq unfree\rq\rq \, in
our actions. The future being a direct outcome of the past, our life is nothing but a series of events each completely determined by its
predeccessors. This was a paradox and irony that has ever since haunted the modern epoch. To overcome this dilemma, philosophers assumed
a kind of \lq\lq duality\rq\rq \, in man; the existence of \lq\lq consciousness\rq\rq \, outside nature. In that sense, the mind was governed by some
mystical laws transcending the mechanistic universe. A separation between mind and matter was thus introduced, which remained ever since,
expressing itself in different philosophical doctrines. This line of demarcation was never crossed: the \lq\lq materialistic\rq\rq
\,world on one side (mind is a by-product of matter), the \lq\lq idealistic\rq\rq \, world on the other (mind precedes matter). Quantum theory abolishes this completely.
Not only that the future is not an outcome of the present, but that the present itself is not completely
specified or determined. Assuming that the human mind operates in accordance to quantum, that is, probabilistic laws,
there
will always be an element of \lq\lq unpredectability\rq\rq \, and \lq\lq novelty\rq\rq\, in its outcome. In other words,
{\it intelligence cannot be tamed}!\footnote{Godel's
undecidability gives a somewhat similar conclusion. Godel's result: the consistency of arithmetic cannot be established
by any metamathematical reasonning which can be represented within the formalism of arithmetic. Consequently, the concept of mathematical
truth cannot be encapsulated in any
formalistic scheme. Mathematical truth is something that goes beyond mere formalism. Stated differently, one can say that
the essence of Godel's result is the following: Human intelligence {\bf transcends} mere formalism and mechanicizability. Again there is always an element of
\lq\lq unexpectedness\rq\rq in the outcome of the human mind!}

\bigskip

We end the paper with the following philosophical debate between Albert Einstein and Niels Bohr.

\subsection{A debate between Einstein and Bohr}

Bohr: Quantum mechanics seems to accord a special role to an observer who is outside the system under study. The information that we as observers
have about the quantum system is coded into the construction of the quantum state of the system. This is necessarily an abstract concept, not
for something in nature, but for a mathematical entity that is invented to keep track of the information that one part of the universe can have about another
part. The quantum state is not a property of the system it describes. It is the property of the boundary or the interface that separates that system from
the rest of the universe, including the observer who studies it. Since a quantum state changes when we make a measurment, I think that it is nothing
but an encoding of what we know.

\bigskip

Einstein: I agree with you that the abstract state space used to represent the information we have is not necessarily something that is in complete
correspondance with the system. If we ask a new question and gain a new information, we will represent it by a new state. This abrupt change is a reflection
of change in our knowledge of the system and not a change in the system itself. However, the fact that it seems intrinsically impossible for one observer to
have all the information that would be necessary to give a complete description of the world seems to me unacceptable. Heisenberg uncertainty
says that
we are allowed to know at most half the information that would be necessary to fully describe any physical system. I think that there is
some crucial element missing in the
\lq\lq Heisenberg, Shrodinger, Dirac \rq\rq formulation of the quantum theory as we understand it today. Nowadays, every Tom, Dick and Harry think they
understand quantum theory. I think they are mistaken. I believe that the goal of physics is to
construct a description of the world as it would be in our absence.

\bigskip

Bohr: I agree with you that in quantum mechanics we deal not only with a description of the system itself, but what we can know about it. We envision a
situation in which the world is divided into two parts. On one side is the particular system under study. On the other side, are ourselves, as observers,
whatever tools and instruments we intend to use in the study. This is very different from the description of the world in the classical context. There
we are invited to imagine that mathematics gives a picture of reality in which the observer need not be glimpsed. While it is true that
quantum theory
doesn't provide an objective picture of the world, this may actually be the great virtue of the theory. It frees us from the fiction of absolute
observer, looking at everything from outside the world. We may miss the picture of the world given to us by classical physics. However, this
idea of representing the whole universe as a collection of classical trajectories reflects a fictitious ideal that corresponds only very approximately to
what is real.

\bigskip

Einstein: I fail to understand your subtle reasoning. You seem to suggest that the electron is a ghost. It is not here, it is not there, it is in a
\lq\lq state that is some mixture of here and there\rq\rq. According to your view, only states and not transition of states is what makes sense in the
context of the micro. I don't like this fuzziness. You are basically saying that there is no world out there. There is only the abstract quantum physical description.
The task of physics is not to find out what nature is. According to your view, physics is concerned with
what we can say about nature and not nature itself. I simply
don't agree. If probability is the language of our world, then the problem is in the quantum theory and not in the quantum world. The wavefunction does
not provide a complete description of physical reality. However, I do believe that we should leave open the question of whether or not such a description
exists.

\bigskip

Bohr: I don't see why you reject quantum theory. I believe that the logic of the micro world (and not the quantum theory alone) is bizzare. What
quantum theory seems to tell us is the following: There is no such thing as an objective reality. Reality (whatever this means) reveals itself through our
methods of questioning. The electron in one context (experimental framework) may act like a particle and in another context it may act like a wave. The
essence of the electron is the sum total of al its manifestations (revealed in different contexts). It is not only a wave, it is not only a particle. It is in
some sense both and neither. We actually create the required property by choosing the suitable experimental context.

\bigskip

Einstein: The EPR thought experiment \footnote{The Einstein - Podelski - Rosen experiment (EPR for short) seems to suggest that at the quantum level, processes are involved
that are somehow controlled at a {\bf global} rather than the {\bf local} scale of things. The remarkable piece of work carried out by the
Irish physicist John Bell, in the early 1960's proved that quantum theory is essentially {\bf non-local}. What Bell did was to find a way to
test directly the principle of locality. Bell found that, in certain cases, the predictions of any local theory must satisfy certain constraints, namely,
the Bell inequalities. Quantum theory, being non-local, violates these constraints!} seems to suggest that quantum
theory is somewhat incompatible with the central idea in special theory of
relativity, namely that of causality. This notion of \lq\lq action at a distance\rq\rq\, makes me
feel uneasy about the quantum
theory. A change here producing
"instantanuously" an affect there is too much for me to absorb. This means, as far that I understand,
that quantum theory is a non-local theory.
The property of {\bf entanglment}\footnote{Whenever two systems interacted, it is more common to find them sharing properties in this way, that to find them in states
such that each have definite individual properties. Quantum theory says quite generally that whenever two
systems have interacted,
their description is tied together in this way; \lq\lq entangled\rq\rq,\, no matter how far apart they may be.}
- every object of the universe being {\it coupled} to every other object, which is the key idea in my
general theory of relativity,
seems to be an essential feature of the quantum world as well. We thus need a mathematical
formulation of such a {\it global} notion. I still believe, I might be wrong
of course, that {\it Heisenberg's {uncertainty principle} may be an approximation to a more
general deterministic law
provided that the notion of entanglment finds a clear
unambiguous mathematical expression}. In other words, a clear mathematical description of the notion of
entanglment may possibly lead to the elimination of the probabilistic and fuzzy laws upon which the quantum theory
is based which I still believe is not a genuine or a true reflection of how the micro-world operates. Accordingly,
we would obtain a new formulation of the quantum theory based on exact non-probabilistic laws.
In this more general theory,  I imagine that a new kind of physics may emerge in which the world is described as
a single entity so that the world around us no longer consists of
a large number of autonomous atoms the properties of which owing nothing to the others. {\it Instead, the
world would be described as a vast, interconnected system of relations in such a way so that the
properties of a single elementary particle or the identity of a point in space requires and
reflects the whole rest of the universe.}

\bigskip

Bohr: I agree with you that we do need a mathematical formulation of the notion of
entaglment or quantum correlation (as
it is technically known).
However, I don't see that if we do succed in achieving this goal, then this may imply that the
\lq\lq uncertainty principle is an approximation to a more general deterministic law\rq\rq, to
use you own words. In fact, it is possible (and most probable) that the notion of entanglment turns
out to be compatible with (if not a consequence of) the uncertainty relation. The \lq\lq new physics\rq\rq
\,emerging in the context of this new theory (still based on the uncertaintity principle), I imagine, should be
interpreted as follows: No single
observer can have complete knowlegde of the world. This is basically what the
uncertainty principle tells us. Accordingly, a {\it complete} description of the universe is only possible
from the point of view of {\it many} observers: Briefly, \lq\lq {\bf I} {\it cannot know everything}, but {\bf we}, in
principle, {\it can know everything}\rq\rq.\, We thus should have a large set of \lq\lq quantum states\rq\rq, each
one of which describes the {\it partial} knowledge that an observer has about the other
things in the universe including his knowledge of the information the other observers hold about the universe, without
actually knowing the content of this information. All this is compatible with the limitations imposed by
the uncertainty relation. The new theory, taking into account all
possible views, must rely on some general principle that
{\it constrains} how the different views may \lq\lq alter\rq\rq,\, while still being partial views of the \lq\lq
same world\rq\rq. The \lq\lq compatibility\rq\rq of the different views; that is, the necessity that
the views of the different observers should be in \lq\lq harmony\rq\rq with each other; cohere, so to speak, would,
I, conjecture, be a natural consequence of the sought for mathematization of the notion of entaglment. We again
arrive at a \lq\lq wholistic
quantized theory\rq\rq\, of the universe. This wholistic theory would be a generalization of
our established quantum theory, though still having the uncertaintity principle as its logical core.
This again reflects my belief that a physical theory does not describe reality independently of our existence,
but is merely a description of our relation with the external world.
The property of every object is both a result and a reflection of its interaction with the rest of the cosmos.

\bigskip

Einstein: I certainly don't agree with  David Bohm's hidden variable theory. I think his idea of a
\lq\lq pilot wave\rq\rq is even more absurd than
the notion of action at a distance. However, I do see some relevence in his \lq\lq implicate order
interpretation\,\rq\rq of the quantum theory.\footnote{Implicit order is a term coined by the physicist
David Bohm to describe the sort of unfolded order that is characteristic of the quantum theory. It is
to be contrasted with the explicit orders of Newtonian mechanics. Bohm believed that this
implicate order has a universal importance and might be useful in understanding the nature of
conciousness.} I
think that your Copenhagen interpretation - this {\it dicotomy} between subject and object -
is not applicable if our \lq\lq laborotary\,\rq\rq
is the whole universe. There is simply {\bf no} place for an external observer. It is in this sense that the
uncertainty principle, the corner stone of the
\lq\lq conventional\rq\rq\, quatum {theory}, may be
an approximation to a yet undiscovered hidden exact law that would establish my conviction that laws of physics
transcend our own existence. A cosmological quantum theory, that is, a quantum theory of the
{\it whole} universe, will necessarily
{\it abolish} the idea of an external observer upon which your Copenhagenhagen interpretation is based.
Such a seperation is possible as long as the system under study is a {\it proper}
part of the universe. Obviously, any theory whose domain is the entire
universe cannot make such a distinction; the line
of demarcation between the subject (observer) and object (observed) will inevitably dissapear. Then
there might evolve a new order of reality; a {\it deterministic non-probabilistic theory that reflects
this smooth merging of the observed and the observer}. Our conventional
quantum theory would then be an approximation of such a theory when restricted to any {\it proper} part of the universe.

\end{mysect}

\bibliographystyle{plain}

\end{document}